\documentclass{article}

\usepackage{arxiv}

\usepackage[utf8]{inputenc} 
\usepackage[T1]{fontenc}    
\usepackage[hypertexnames=false]{hyperref}       
\usepackage{url}            
\usepackage{booktabs}       
\usepackage{amsfonts}       
\usepackage{nicefrac}       
\usepackage{microtype}      
\usepackage{graphicx}
\usepackage[numbers]{natbib}
\usepackage{doi}
\usepackage{bm}
\usepackage{amsmath}
\usepackage{xcolor}
\usepackage{comment}
\usepackage{multirow}
\usepackage{float}
\usepackage{makecell}
\usepackage{amssymb}
\title{Unraveling time-varying causal effects of multiple exposures: integrating Functional Data Analysis with Multivariable Mendelian Randomization}

\author{
\fontsize{9.7}{11}\selectfont
Nicole Fontana$^{1,2}$,
Francesca Ieva$^{1,2}$,
Luisa Zuccolo$^{2}$,
Emanuele Di Angelantonio$^{2,3,4,5,6,7}$,
Piercesare Secchi$^{1}$
}

\date{
$^{1}$ MOX, Department of Mathematics, Politecnico di Milano, Milan, Italy\\
$^{2}$ Health Data Science Research Centre, Human Technopole, Milan, Italy\\
$^{3}$ British Heart Foundation Cardiovascular Epidemiology Unit, Department of Public Health and Primary Care, University of Cambridge, Cambridge, UK\\
$^{4}$ Victor Phillip Dahdaleh Heart and Lung Research Institute, University of Cambridge, Cambridge, UK\\
$^{5}$ British Heart Foundation Centre of Research Excellence, University of Cambridge, Cambridge, UK\\
$^{6}$ National Institute for Health and Care Research Blood and Transplant Research Unit in Donor Health and Behaviour, University of Cambridge, Cambridge, UK\\
$^{7}$ Health Data Research UK Cambridge, Wellcome Genome Campus and University of Cambridge, Cambridge, UK\\[6pt]
\texttt{nicole.fontana@polimi.it}
}

\renewcommand{\shorttitle}{}

\hypersetup{
pdftitle={Unraveling time-varying causal effects of multiple exposures: integrating Functional Data Analysis with Multivariable Mendelian Randomization},
pdfauthor={Nicole Fontana, Francesca Ieva, Luisa Zuccolo, Emanuele Di Angelantonio, Piercesare Secchi},
pdfkeywords={},
}

\begin{document}
\maketitle

\begin{abstract}
Mendelian Randomization is a widely used instrumental variable method for assessing causal effects of lifelong exposures on health outcomes. Many exposures, however, have causal effects that vary across the life course and often influence outcomes jointly with other exposures or indirectly through mediating pathways. Existing approaches to multivariable Mendelian Randomization assume constant effects over time and therefore fail to capture these dynamic relationships. We introduce Multivariable Functional Mendelian Randomization (MV-FMR), a new framework that extends functional Mendelian Randomization to simultaneously model multiple time-varying exposures. The method combines functional principal component analysis with a data-driven cross-validation strategy for basis selection and accounts for overlapping instruments and mediation effects. Through extensive simulations, we assessed MV-FMR’s ability to recover time-varying causal effects under a range of data-generating scenarios and compared the performance of joint versus separate exposure effect estimation strategies. Across scenarios involving nonlinear effects, horizontal pleiotropy, mediation, and sparse data, MV-FMR recovered the underlying functional effect patterns and outperformed univariable approaches. To demonstrate its practical utility in a proof-of-concept application, we applied MV-FMR to UK Biobank data to investigate the time-varying causal effects of systolic blood pressure and body mass index on coronary artery disease. MV-FMR provides a flexible and interpretable framework for disentangling complex time-dependent causal processes and offers new opportunities for identifying life-course critical periods and actionable drivers relevant to disease prevention.
\end{abstract}


\section{Introduction}\label{sec:intro}
Establishing causal relationships between exposures and health outcomes remains a central challenge in epidemiological research. Mendelian Randomization (MR) is a widely used method for causal inference that leverages genetic variants as instrumental variables (IVs) to estimate the causal effect of an exposure on an outcome \cite{sanderson2022mendelian, Lawlor2008}. Traditional MR methods generally assume that a single exposure exerts a constant causal effect. Since genetic variants are fixed at conception, estimates from traditional MR are typically interpreted as reflecting the average effect of genetically predicted exposure across the lifetime, thereby capturing cumulative rather than age-specific or period-specific influences~\cite{burgess2021interpretation}. However, many exposures exhibit time-varying effects, meaning that their causal impact on outcomes can change across different stages of life~\cite{Lee2022AgeDependentBMI}. For example, the effect of body mass index on cardiovascular disease risk may differ between early adulthood and older age. This variation can occur due to evolving physiological processes or age-specific biological mechanisms~\cite{Palmer2022, Karlsson2020}. To capture these dynamics, analytical methods must move beyond single time-point estimates and instead model how effects evolve over time. This approach allows for the identification of age-specific or life course-specific influences that may be overlooked when relying on a single measurement of exposures~\cite{wagner2021time, shi2022mr_timevarying}. 

Moreover, emerging evidence indicates that genetic associations with certain exposures can vary across the life course within a population, meaning that MR estimates may be more informative about exposure effects during particular life periods rather than about a uniform effect over time~\cite{sanderson2022timevarying, jiang2021age}. Recent work has further highlighted the empirical relevance of this issue by developing a framework for testing age-varying associations between polygenic risk scores and phenotypes, reporting widespread age variation in genetic associations across many complex traits~\citep{yang2025taps}. Although this work focuses on the temporal pattern of genetic association with a trait rather than the causal effect of a time-varying exposure on an outcome, these findings support the importance of accounting for time-varying genetic effects in MR modelling. Incorporating such temporal variation into MR analyses can help identify life periods during which an exposure may have greater impact on health outcomes~\cite{Power2024Mendelian}, thereby informing the design and timing of preventive interventions.

In response to this challenge, recent work has focused on estimating time-varying causal effects using MR. Several approaches have represented exposure measurements at different time points as separate variables, thereby discretizing time and only partially exploiting the continuous nature of individual exposure trajectories~\cite{cao2016timevarying, richardson2020earlylater, shi2022mr_timevarying, sanderson2022timevarying, richardson2023timevarying}. Tian et al.~\cite{Tian2024} proposed a continuous-time MR framework, called Univariable Functional Mendelian Randomization (U-FMR), which models the exposure effect as a smooth function of time using functional data analysis~\cite{ramsay2005fda}. Their method uses Functional Principal Component Analysis (FPCA)~\citep{Yao2005} to reduce the dimensionality of longitudinal exposure data and integrates these low-dimensional representations into an MR model. This framework provides a flexible and theoretically robust approach for inferring how causal effects evolve over the life course, given longitudinal data. Related work has incorporated longitudinal exposure information into multivariable MR by summarizing individual trajectories with linear mixed models and using the resulting features to estimate the causal effects of average exposure level, change over time, and within-subject variability~\citep{pott2026longitudinal}. This provides a computationally tractable way to use repeated exposure measurements, but represents the exposure history through a finite set of summaries rather than estimating a causal effect function defined continuously over time.

Despite these advances, existing MR methods that investigate the life course effects of exposure have focused exclusively on single exposures. However, in real-world scenarios, multiple exposures often act simultaneously, share genetic predictors, or influence each other over the life course, particularly when they are biologically related. Multivariable MR (MV-MR) model extends the univariable MR (UVMR) framework to jointly estimate the causal effects of multiple exposures using a set of genetic variants as instruments~\cite{sanderson2019examination}. Unlike the UVMR model, which captures only the total effect of a single exposure, MV-MR model can disentangle the direct effects of many exposures, provided that the genetic instruments are sufficiently associated with each exposure throughout the life course. However, standard MV-MR methods treat each exposure as time-invariant and therefore cannot estimate how the direct effect of each exposure varies with age. 

To address the limitation of time-invariant effects, we propose Multivariable Functional Mendelian Randomization (MV-FMR), which extends U-FMR~\cite{Tian2024} to simultaneously model multiple time-varying exposures. Our key methodological contributions include: (1) a data-driven cross-validation procedure for selecting the number of basis components for each exposure, (2) joint modelling of time-varying exposures that share common genetic pathways, (3) handling of mediation pathways between time-varying exposures, and (4) implementation for both continuous and binary outcomes. 
To evaluate the performance of MV-FMR, we conducted extensive simulation studies under three main scenarios: horizontal pleiotropy through overlapping genetic instruments, mediation pathways where one exposure influences another, and null effects to assess specificity. We further performed sensitivity analyses examining different functional forms for the exposure-outcome effects (linear and nonlinear effects) and outcome types (continuous and binary). Results showed that MV-FMR recovered the underlying time-varying causal effect patterns and outperformed the corresponding univariable functional MR approaches under the simulation settings considered.
We apply MV-FMR to UK Biobank data in a proof-of-concept analysis to estimate the time-varying causal effects of systolic blood pressure and body mass index on coronary artery disease risk across adulthood, demonstrating how the method can identify critical age periods and potentially disentangle direct effects from mediation pathways that would be missed by traditional approaches.

The remainder of the paper is organized as follows. In Section~\ref{sec:method}, we review the background on MV-MR and U-FMR models, and introduce our proposed extension, including its statistical formulation and estimation strategy. Section~\ref{sec:simulation} presents simulation studies evaluating the performance of MV-FMR compared to the univariable approach. In Section~\ref{sec:application}, we apply the method in a proof-of-concept application using UK Biobank data. Finally, Section~\ref{sec:discussion} concludes with a discussion of the results, limitations, and directions for future research.

\section{Methods}
\label{sec:method}
\subsection{Background on Multivariable Mendelian Randomization (MV-MR)}\label{subsec2.1}
Multivariable MR (MV-MR) model generalizes the classical MR framework to jointly estimate the direct causal effects of multiple exposures on an outcome using genetic variants as instrumental variables~\cite{sanderson2021multivariable}. While univariable MR captures the total effect of a single exposure, MV-MR model disentangles the direct effect of each exposure while accounting for the influence of the others. Estimating multiple exposures simultaneously is often desirable, for example, when exposures are biologically or clinically related, or when one exposure may mediate the effect of another on the outcome. By modelling multiple exposures simultaneously, MV-MR model can disentangle overlapping causal mechanisms and improve robustness to pleiotropy, provided that pleiotropic effects operate only through the modelled exposures. The validity of the model requires an extension of the standard IV assumptions to the multivariable setting:
\begin{enumerate}
    \item Relevance (MV-IV1): each instrument must be strongly associated with at least one exposure, conditional on the others.
    \item Exclusion Restriction (MV-IV2): instruments must affect the outcome only through the included exposures. 
    \item Independence (MV-IV3): each instrument must be independent of any confounders of the exposures-outcome relationship.
\end{enumerate}
These assumptions follow the standard identification conditions for MV-MR models~\citep{sanderson2019examination,sanderson2021multivariable}.

\subsection{Overview of the Univariable Functional Mendelian Randomization (U-FMR) with continuous outcome}\label{subsec2.2}
We begin by briefly showing the U-FMR model proposed by~\cite{Tian2024} as a foundation for our multivariable extension presented in Section~\ref{subsec2.3}. The setting involves a continuous outcome $Y$, a time-varying exposure $X(t)$, and a set of $P$ genetic instruments $\bm{G}$. Inference on time-varying causal effects relies on life-course variation in the associations between the genetic instruments and the exposure. When such associations evolve over time, they provide information that enables the identification of how the causal effect of the exposure on the outcome changes across the life course. Under this setting, the instruments act as valid proxies for the exposure across the observation period, enabling the estimation of the time-varying causal effect. The causal effect is therefore modelled as a smooth function $\beta(t)$, capturing how the impact of $X(t)$ on $Y$ evolves over the observed time interval. This leads to the structural model:
\begin{equation}\label{eq:1}
    Y = \beta_0 + \int_0^T \beta(t) X(t) \, dt + \varepsilon.
\end{equation}

Because exposure measurements are often sparse and noisy, individual trajectories $X(t)$ are treated as functional objects and modelled in a linear additive form using a basis expansion derived from FPCA:
\begin{equation}\label{eq:2}
    X(t) \approx \mu(t) + \sum_{k=1}^{K} \xi_k \phi_k(t),
\end{equation}
where $K$ denotes the number of selected principal components, $\mu(t)$ is the mean exposure trajectory, $\phi_k(t)$ are eigenfunctions of the covariance operator, and $\xi_k$ are subject-specific FPCA scores. Similarly, the causal effect function $\beta(t)$ is represented using as basis functions the eigenfunctions of the FPCA:
\begin{equation}\label{eq:3}
    \beta(t) = \sum_{k=1}^{K} \alpha_k \phi_k(t),
\end{equation}
where $\alpha_k$ are the corresponding coefficients to be estimated.
Substituting the expansions for $X(t)$ and $\beta(t)$ from equations~\eqref{eq:2} and~\eqref{eq:3} into the structural model~\eqref{eq:1}, the equation simplifies to:
\begin{equation}\label{eq:4}
    Y \approx \beta_0^* + \sum_{k=1}^{K} \beta^*_k \xi_k,
\end{equation}
where $\beta_0^* = \beta_0+\int_0^T\beta(t)\mu(t)dt$ and each pseudo-exposure coefficient $\beta^*_k$ is defined as:
\begin{equation}
    \beta^*_k = \int_0^T \beta(t) \phi_k(t) \, dt.
\end{equation}
This formulation reduces the original functional model to a standard linear regression on the subject-specific FPCA scores $\xi_k$, making it more tractable for estimation and inference in settings with sparse or noisy exposure measurements.

The FPCA scores $\xi_k$ are treated as pseudo-exposures, and their corresponding causal coefficients $\beta^*_k$ are estimated using the continuously updating Generalized Method of Moments (GMM), where genetic variants $\bm{G}$ are used as instruments for the FPCA scores $\xi_k$. The continuously updating GMM estimator is used here because it has been shown to have smaller finite-sample bias than alternative GMM estimators~\cite{hansen1996finite}.
All variables are mean-centered prior to estimation to ensure numerical stability. The sample moment function is defined as:
\begin{equation}
\mathbf{g}(\boldsymbol{\beta}^*) = \frac{1}{n}\mathbf{G}^T(\mathbf{Y} - \boldsymbol{\xi}\boldsymbol{\beta}^*),
\end{equation}
and the GMM estimator is obtained by minimizing the continuously updated objective function~\cite{hansen1996finite}:
\begin{equation}\label{eq:cugmm_objective}
\boldsymbol{\hat{\beta}}^* = \arg\min_{\boldsymbol{\beta}^*} \mathbf{g}(\boldsymbol{\beta}^*)^T \mathbf{W}(\boldsymbol{\beta}^*)^{-1} \mathbf{g}(\boldsymbol{\beta}^*),
\end{equation}
where the weighting matrix is $\mathbf{W}(\boldsymbol{\beta}^*) = \frac{1}{N}\mathbf{G}^T \text{diag}(\mathbf{Y} - \boldsymbol{\xi}\boldsymbol{\beta}^*)^2 \mathbf{G}$. Under standard regularity conditions, $\boldsymbol{\hat{\beta}}^*$ is consistent and asymptotically normally distributed. 

Once the coefficients $\beta^*_k$ are estimated, the functional effect $\beta(t)$ can be reconstructed by inverting the relationship between $\beta(t)$ and the FPCA basis using:
\begin{equation}
\beta(t) \approx \sum_{k=1}^K \beta^*_k \phi_k(t).
\end{equation}

\begin{figure}[h!] 
\centering
\includegraphics[width=0.8\textwidth]{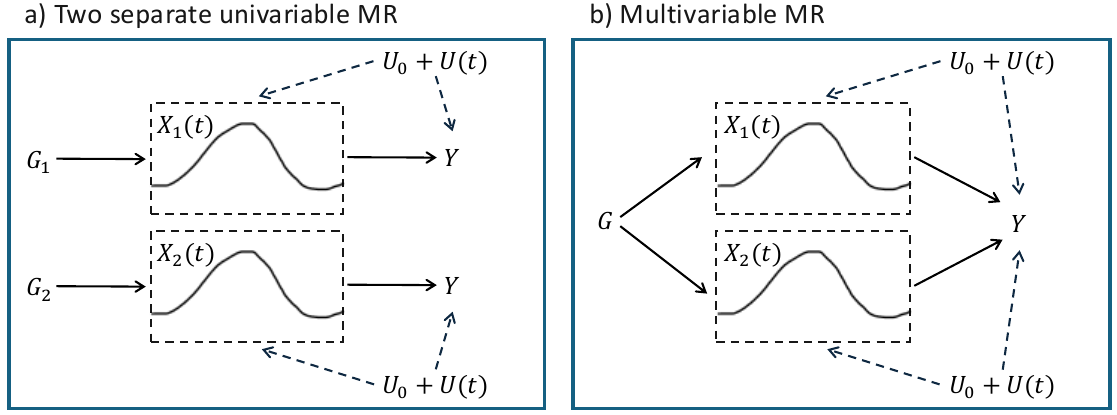} 
\caption{Comparison of (a) Univariable functional MR (independent effects of the exposures) with (b) Multivariable functional MR (joint effects of the exposures). The diagram illustrates the case of two exposures for simplicity, although the framework can be generalized to more than two exposures. Dashed arrows represent confounding pathways.}
\label{fig:comparison_MR} 
\end{figure}

This framework provides the basis for extending the model to multiple exposures, as illustrated in Figure~\ref{fig:comparison_MR} for a simple setting with two exposures. While U-FMR (panel a) estimates total effects separately, MV-FMR (panel b) jointly models both exposures to recover direct causal effects.

\subsection{Multivariable Functional Mendelian Randomization (MV-FMR) with continuous outcome}\label{subsec2.3}
Building on the U-FMR framework described in Section~\ref{subsec2.2} and the causal structure illustrated in Figure~\ref{fig:comparison_MR}, we extend the model to a multivariate setting to accommodate multiple time-varying exposures, each with its own dynamic causal effect. Consider $m \geq 2$ exposures $X_1(t), \ldots, X_m(t)$, observed over $t \in [0, T]$. We assume that each exposure trajectory can be represented using FPCA:
\begin{equation}~\label{eq:multi_fpca}
X_{j}(t) = \mu_j(t) + \sum_{k=1}^{K_j} \xi_{j,k} \, \phi_{j,k}(t) + \epsilon_j(t), \quad j = 1, \ldots, m,
\end{equation}
where $\mu_j(t)$ is the mean trajectory, $\phi_{j,k}(t)$ are eigenfunctions and $\xi_{j,k}$ are FPCA scores, for exposure $j$. The outcome is modelled as a function of all $m$ exposures:
\begin{equation}\label{eq:multi_structural}
Y = \beta_0 + \sum_{j=1}^{m} \int_0^T \beta_j(t) X_{j}(t) \, dt + g_Y(U, \epsilon_{Y}),
\end{equation}
where $\beta_j(t)$ denotes the direct time-varying causal effect of exposure $j$, and $g_Y(U, \epsilon_{Y})$ captures unmeasured confounding and random noise. By substituting the FPCA expansion~\eqref{eq:multi_fpca} into~\eqref{eq:multi_structural}, we can rewrite the model as:
{\footnotesize
\begin{equation}\label{eq:model}
\begin{aligned}
Y &= \beta_0 + \sum_{j=1}^{m} \int_0^T \beta_j(t) \left[ \mu_j(t) + \sum_{k=1}^{K_j} \xi_{j,k} \, \phi_{j,k}(t) + \epsilon_{j}(t) \right] dt + g_Y(U, \epsilon_{Y}) \\
&= \underbrace{\beta_0 + \sum_{j=1}^{m} \int_0^T \beta_j(t) \mu_j(t) \, dt}_{\beta_0^*} + \sum_{j=1}^{m} \sum_{k=1}^{K_j} \underbrace{\left[ \int_0^T \beta_j(t) \phi_{j,k}(t) \, dt \right]}_{\beta_{j,k}^*} \xi_{j,k} + \tilde{\epsilon},
\end{aligned}
\end{equation}
} where $\tilde{\epsilon} = \sum_{j=1}^m \int_0^T \beta_j(t)\epsilon_j(t)dt + g_Y(U,\epsilon_Y)$ collects the contribution of exposure-specific residual variation not captured by the retained FPCA components, unmeasured confounding, and outcome noise. Under the MV-IV assumptions stated above, genetic variants provide exogenous variation in the FPCA scores that is independent of unmeasured confounding and exposure-specific measurement error. As in conventional MR, instrumenting the FPCA scores with genetic variants can therefore reduce bias arising from these sources, but does not protect against bias due to invalid instruments or model misspecification. This formulation reduces the multivariate functional problem to an MV-MR model, in which the FPCA scores $\xi_{j,k}$ act as pseudo-exposures, and the corresponding coefficients $\beta_{j,k}^*$ can be estimated using genetic instruments. Consequently, the effective dimension of the MV-MR model is determined by the total number of retained FPCA scores, $\sum_{j=1}^{m}K_j$, rather than by the number of original functional exposures alone.

\subsubsection{Estimation of component effects}\label{subsubsec2.3.2}
Each exposure trajectory is represented using FPCA, with scores estimated via the PACE method~\citep{Yao2005} to accommodate sparse and irregular measurements. Full mathematical details of the FPCA representation and PACE estimation are provided in Supplementary Method 1. Let $\mathbf{G} \in \mathbb{R}^{N \times P}$ denote the full matrix of genetic instruments. For intuition, one may view $\mathbf{G}$ as including both variants that are associated with individual exposure trajectories and variants that are shared across multiple exposures. For example, in the two-exposure case, $\mathbf{G} = [\mathbf{G}_1, \mathbf{G}_{12}, \mathbf{G}_2]$, where $\mathbf{G}_1$ and $\mathbf{G}_2$ denote variants primarily associated with $X_1(t)$ and $X_2(t)$, respectively, and $\mathbf{G}_{12}$ denotes variants associated with both exposures. This partition is introduced only for illustration, since for a general number of exposures $m$, the full matrix $\mathbf{G}$ is used directly in the estimation without imposing a rigid separation between exposure-specific and shared instruments. 
The estimation of $\boldsymbol{\beta}^*_1, \ldots, \boldsymbol{\beta}^*_m$ follows the GMM framework described in Section~\ref{subsec2.2}, extending it to jointly model all $m$ exposures. The stacked FPCA scores $\boldsymbol{\xi} = [\boldsymbol{\xi}_1, \ldots, \boldsymbol{\xi}_m] \in \mathbb{R}^{N \times \sum_{j=1}^{m} K_j}$ and coefficient vector $\boldsymbol{\beta}^* = [\boldsymbol{\beta}^*_1, \ldots, \boldsymbol{\beta}^*_m]$ are used in the GMM objective function with $\mathbf{G}$ as the full instrument matrix. This joint estimation allows the model to account for correlation between exposures through overlapping instruments while maintaining the asymptotic properties established for the univariable case.

\subsubsection{Data-driven selection of FPCA components}\label{subsubsec2.3.3}
To approximate each exposure trajectory, the FPCA expansion is truncated at $K_j$ components. The number of retained FPCs $K_j$ critically determines the flexibility of the causal effect estimation. We select $K_j$ using a data-driven k-fold cross-validation procedure that optimizes predictive performance. This approach in this setting outperforms variance-based criteria, which often retain too few components and fail to capture higher-order effects. Model performance is evaluated via mean squared error (MSE). Full details of the data-driven selection algorithm are provided in Supplementary Method 1. This approach balances model complexity and generalizability, capturing meaningful functional effects while minimizing overfitting.

\subsubsection{Reconstructing the coefficient functions} \label{subsubsec2.3.4}
The time-varying causal effect functions $\beta_j(t)$ are reconstructed from the estimated pseudo-exposure coefficients using the eigenfunction basis. Let $\beta_{j,k}^* = \int_0^T \beta_j(t) \phi_{j,k}(t) dt\, $ then the reconstruction formula is given by: 
\begin{equation}
\hat{\beta}_j(t) = \boldsymbol{\phi}^T_j(t) \boldsymbol{\hat{\beta}}^*_j \quad \forall  j = 1, \ldots, m,
\end{equation}
where $\boldsymbol{\phi}_j(t) = [\phi_{j,1}(t), \ldots, \phi_{j,K_j}(t)]^T$ denotes the vector of eigenfunctions and $\boldsymbol{\hat{\beta}}^*_j \in \mathbb{R}^{K_j}$ is the corresponding vector of estimated coefficients. We employ a non-parametric bootstrap procedure for constructing confidence intervals. By resampling individuals with replacement, we construct pointwise 95\% confidence intervals for $\hat{\beta}_j(t)$ using the $2.5^{\text{th}}$ and $97.5^{\text{th}}$ percentiles of the bootstrap distribution at each time point $t$. This procedure captures sampling variability in the estimated causal coefficients and the corresponding reconstructed effect functions. Full details are provided in Supplementary Method 2.

\subsubsection{Two-sample setting}
Although the primary implementation of MV-FMR considered in this work uses a one-sample design, the same estimating framework can also be formulated in a two-sample setting. In this extension, genetic associations with the FPCA-derived exposure scores are estimated in an exposure sample, while genetic associations with the outcome are obtained from an independent outcome sample, typically using summary-level GWAS data. When the two samples are non-overlapping and drawn from comparable populations, this design avoids bias due to sample overlap and enables the use of large-scale GWAS summary statistics for the outcome. Details of the two-sample formulation are provided in Supplementary Method~3.

\subsection{Multivariable Functional Mendelian Randomization (MV-FMR) with binary outcome}\label{subsec2.4}
We also propose an extension of the MV-FMR framework to binary outcomes by modifying the structural equation and estimation procedure while preserving the functional representation of exposures~\citep{fontana2026modelling}. The outcome model in~\ref{eq:multi_structural} becomes:
\begin{equation}\label{eq:multi_structural_binary}
\text{logit}(\mathbb{P}(Y = 1)) = \beta_0 + \sum_{j=1}^{m} \int_0^T \beta_j(t) X_{j}(t) \, dt + g_Y(U).
\end{equation}
The FPCA representation of exposures in~\ref{eq:multi_fpca} and the reduction to pseudo-exposures in~\ref{eq:model} remain unchanged, yielding a logistic model for the FPCA scores.

For binary outcomes, we apply a two-stage residual inclusion (2SRI) control function approach to account for endogeneity in the FPCA scores. 
This method is preferred for nonlinear models over standard two-stage predictor substitution (2SPS) approaches, which replace endogenous variables with their predicted values. In the context of nonlinear models such as logistic regression, the 2SPS estimator produces inconsistent estimates because the nonlinear link function invalidates the substitution of predicted values, thereby distorting the conditional mean structure~\citep{terza2008two, wooldridge2015control}. In contrast, the 2SRI approach controls for endogeneity by including first-stage residuals as additional regressors. Following the notation introduced in Section~\ref{subsec2.3}, in the first stage, FPCA scores are regressed on instruments:
\begin{equation}
\boldsymbol{\xi} = \mathbf{G}\boldsymbol{\Gamma} + \mathbf{V}, \quad \hat{\mathbf{V}} = \boldsymbol{\xi} - \mathbf{G}\hat{\boldsymbol{\Gamma}},
\end{equation}
where $\hat{\mathbf{V}}$ are the residuals capturing variation in FPCA scores not explained by the instruments. The second stage includes both scores and residuals in a logistic regression: 
\begin{equation}
\text{logit}(\mathbb{P}(Y = 1)) =\mathbf{1}_n \boldsymbol{\beta}_0 + \boldsymbol{\xi} \boldsymbol{\beta}^*+ \hat{\mathbf{V}} \boldsymbol{\alpha}  
\end{equation}
where $\boldsymbol{\hat{\beta}^*}$ estimates the direct effect of the FPCA scores on the outcome and $\boldsymbol{\alpha}$ captures the correlation between the FPCA scores and unobserved confounders, correcting for endogeneity. Robust standard errors are used to account for the estimation of $\hat{\mathbf{V}}$ in the first stage. Under standard regularity conditions and correct specification of the first-stage model, $\boldsymbol{\hat{\beta}^*}$ is consistent and asymptotically normal.

The data-driven component selection procedure (\ref{subsubsec2.3.3}) and the reconstruction of $\hat{\beta}_j(t)$ (\ref{subsubsec2.3.4}) follow the same approach as for continuous outcomes, with the exception that the area under the ROC curve (AUC) replaces MSE as the performance metric for cross-validation in the component selection procedure.

The proposed one-sample and two-sample methods are implemented in the R package \texttt{mvfmr}, available from CRAN~\citep{mvfmr_package}.

\section{Simulation Studies}
\label{sec:simulation}
We conducted simulation studies to evaluate the performance of the proposed MV-FMR approach. The simulations assessed the ability of MV-FMR model to recover time-varying causal effects under different data-generating mechanisms and compared joint multivariable estimation with separate univariable analyses of each exposure. The main simulations are presented with two exposures for simplicity and computational feasibility. However, as shown in Section~\ref{sec:method}, the framework naturally extends to more than two exposures, and we include a sensitivity analysis evaluating its performance in settings with additional exposures.

\subsection{Simulation scenarios}\label{subsec3.1}
\begin{figure}[h!] 
\centering
\includegraphics[width=1\textwidth]{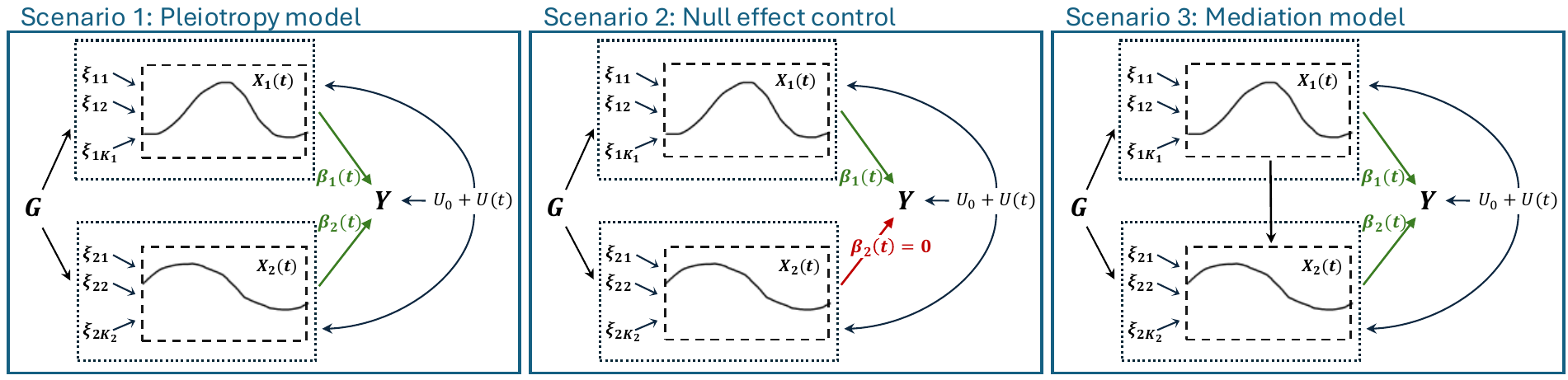} 
\caption{Illustration of the three simulation scenarios used to evaluate the MV-FMR model. Scenario 1: Pleiotropy model (left). Scenario 2: Null effect control (center). Scenario 3: Mediation model (right).}
\label{fig:scenarios} 
\end{figure}

We designed three distinct simulation scenarios, shown in Figure~\ref{fig:scenarios}, to assess the ability of the MV-FMR model to estimate direct time-varying causal effects in the presence of overlapping instruments, null exposure effects, and mediation. Genetic instruments were allowed to be associated with one or both exposures, reflecting the possibility of partially overlapping genetic architecture without requiring a strict distinction between exposure-specific and shared instruments.

\begin{itemize}
    \item \textbf{Scenario 1: Pleiotropy model} (Figure~\ref{fig:scenarios}, left panel): Exposures have an independent pleiotropic pathway from instruments to the outcome. This scenario evaluates the ability of the model to disentangle direct causal effects in the presence of horizontal pleiotropy. 

    \item \textbf{Scenario 2: Null effect control} (Figure~\ref{fig:scenarios}, center panel): Same genetic architecture as in Scenario 1, but the second exposure does not have a causal effect on the outcome $(\beta_2(t) = 0 \forall t)$. This scenario evaluates the pointwise estimation and coverage behaviour of the model when the true causal effect function is null.

    \item \textbf{Scenario 3: Mediation model} (Figure~\ref{fig:scenarios}, right panel): $X_1(t)$ causally influences $X_2(t)$, and both affect the outcome $Y$. This scenario evaluates the ability of the model to recover direct effects while accounting for mediation.
\end{itemize}
    
These three scenarios reflect common epidemiological challenges that arise when investigating multiple correlated risk factors~\cite{sanderson2022timevarying}. Horizontal pleiotropy is frequently observed when genetic variants influence multiple physiological pathways, potentially violating the exclusion restriction in univariable MR analysis. Mediation pathways are pervasive in chronic disease epidemiology, where one risk factor operates both directly on the outcome and indirectly through downstream exposures. In such cases, U-FMR estimates the total causal effect, whereas MV-FMR recovers the direct effect by conditioning on the mediator. Finally, it is important to assess specificity in settings where some exposures have true null effects. In practice, analyses of joint time-varying exposures may include variables that are correlated with causal risk factors but do not themselves affect disease risk, and we aim to verify that the model does not introduce spurious non-null estimates. Correctly distinguishing these relationships is essential for identifying modifiable targets for intervention and avoiding spurious causal inferences.

\subsection{Data generating process}\label{subsec3.2}
\label{subsec:dgp}
We simulate individual-level functional trajectories $X_j(t)$ for each exposure $j = 1, 2$ over the time domain $t \in [0, 50]$, according to the following exposure model:
\begin{equation}
X_{j}(t) = \sum_{l=1}^{P} \alpha_{l}(t) G_{l} + [U_{0j} + U_j(t)] + \epsilon_{j}(t),
\end{equation}
where $G_{l} \sim \text{Binomial}(2, 0.3)$ denotes the $l$-th genetic variant. The time-varying genetic effect $\alpha_{l}(t)$ determines the influence of each variant on exposure $j$ over the observation period, while $U_{0j}$ and $U_j(t)$ denote time-fixed and time-varying confounders affecting the exposure. The residual term $\epsilon_{j}(t)$ accounts for exposure-specific noise and measurement error.
Both $U_j(t)$ and $\epsilon_{j}(t)$ are modelled as discretized Wiener processes, and
$U_{0j} \sim \mathcal{N}(0, 1)$. 
To emulate sparse longitudinal data, for each exposure, each individual is observed at $n_{\text{sparse}}=10$ randomly selected time points over $[0, 50]$. The time-varying genetic effect on the exposures $\alpha_{l}(t)$ is generated:
\begin{equation}
\alpha_{l}(t) = a_{l} + b_{l} \cdot t, \quad a_{l} \sim \text{Uniform}(-0.1, 0.1), \quad b_{l} \sim \text{Uniform}(-0.004, 0.004).
\end{equation}
Instrument strength is calibrated such that the pointwise heritability $h^2(t)$, defined as the proportion of exposure variance attributable to genetic variants at each age $t$, varies between approximately 0.055 and 0.104 over the simulation domain $[0, T]$, with a time-averaged value of 0.071. This value is consistent with the heritability levels commonly used in MR simulation studies for polygenic complex traits. The derivation of $h^2(t)$ in the functional exposure setting, together with the analytical calculation for the simulation design used here, is provided in Supplementary Method~4.

Only for Scenario 3, $X_1(t)$ influences $X_2(t)$ and then is generated as:
\begin{equation}
X_{2}(t) = \sum_{l=1}^{P} \alpha_{l}(t) G_{l} + \gamma \cdot X_{1}(t) + [U_{0j} + U_j(t)] + \epsilon_{2}(t),
\end{equation}
where $\gamma$ controls the strength of the mediation pathway.

The outcome $Y_i$ is simulated according to:
\begin{equation}
Y_i = \beta_0 + \int_0^T \beta_1(t) X_{i1}(t) dt + \int_0^T \beta_2(t) X_{i2}(t) dt + \varepsilon_{Y_i},\varepsilon_{Y_i} \sim \mathcal{N}(0, 1),
\end{equation}
with integration approximated numerically over a regular grid. Three functional forms for the exposure-outcome effects $\beta_j(t)$ are considered:
\begin{itemize}
    \item \textbf{Null effect:} $\beta_j(t) = 0$,
    \item \textbf{Linear effect:} $\beta_j(t) = 0.02 \cdot t$,
    \item \textbf{Quadratic effect:} $\beta_j(t) = 0.002 \cdot t^2 - 0.11 \cdot t + 0.5$.
\end{itemize}

We compare the finite-sample behaviour of MV-FMR with U-FMR applied separately to each exposure. The ability of each method to recover the true time-varying causal effects is assessed using several metrics. The Integrated Squared Error (ISE) quantifies the average squared deviation between the estimated and true time-varying effect functions across time. To distinguish systematic bias from random variation across simulation replications, we also calculate the Integrated Squared Bias (ISB), defined as the integrated squared difference between the mean estimated effect across replications and the true effect function. The pointwise 95\% coverage rate is defined as the proportion of time points at which the 95\% bootstrap confidence interval contains the true effect. Standard error calibration is assessed using the standard error calibration ratio (SEr), defined as the ratio between the average model-based standard error across simulation replications and the empirical standard deviation of the estimates. Values close to one indicate well-calibrated standard errors, values below one indicate underestimated uncertainty, and values above one indicate conservative standard errors. Instrument strength is measured using the conditional F-statistic for each principal component~\cite{Sanderson2021-Fstat}.

Each simulation scenario is replicated $R=500$ times with a sample size $N=5000$. The baseline design uses $P=50$ instruments per exposure, of which $\lfloor p_{12}\cdot P \rfloor$ are shared between the two exposures, with $p_{12}=0.15$ in the baseline simulations. In Scenario~3, the baseline mediation parameter is set to $\gamma=0.30$. Additional simulations vary the degree of instrument overlap across exposures, with $p_{12} \in \{0.15,0.30,0.50,0.80\}$, and the mediation strength in Scenario~3, with $\gamma \in \{0.15,0.30,0.50,0.80\}$, to assess sensitivity to these design parameters. We also assess sensitivity to longitudinal data sparsity by varying the number of observation points per individual exposure, with $n_{\text{sparse}} \in \{5,10,25,50\}$, over the interval $[0,50]$. Further sensitivity analyses vary the total number of instruments, $P \in \{25,50,75,100\}$, to assess the impact of instrument count on estimation accuracy and to characterize the associated heritability. Finally, we conduct a sensitivity analysis varying the number of exposures, with $m \in \{2,3,4,5\}$, to assess the generalizability of the method to settings with more than two time-varying exposures. 

\subsection{Simulation results}\label{subsec3.3}
We first evaluated the performance of MV-FMR across the three scenarios in Figure~\ref{fig:scenarios}, assuming linear time-varying exposure-outcome effects for both exposures. Table~\ref{tab:comparison_continuous} reports the ISE, ISB, pointwise 95\% coverage rate, and mean SE calibration ratio for the MV-FMR model across scenarios, together with the corresponding results obtained by applying U-FMR separately to each exposure. Conditional F-statistics and results for binary outcomes are provided in Supplementary Tables~1 and~2. Overall, MV-FMR recovered the true functional effect shapes well across the scenarios considered, with low ISE and ISB; however, pointwise coverage and standard error calibration varied across scenarios and outcome types. In Scenario~1, MV-FMR showed stable performance in the presence of overlapping genetic variants, with low estimation error across the configurations considered. In Scenario~2, where $\beta_2(t)=0$ by design, MV-FMR correctly recovered the null effect for the second exposure. This indicates appropriate pointwise behaviour under the null function in this setting, with coverage for $\beta_2(t)$ close to the nominal level and ISE close to zero. By contrast, applying the U-FMR model separately to each exposure produced inflated estimates for the null exposure, reflecting its inability to separate the direct effect of each exposure from shared genetic pathways. In Scenario~3, MV-FMR distinguished direct causal effects from mediated pathways for both exposures, whereas the univariable approach conflated these effects, resulting in higher ISE and ISB, lower coverage, and standard error calibration ratios further from 1. As expected, the proposed model showed better estimation and uncertainty calibration for continuous outcomes than for binary outcomes, where the reduced information content of the endpoint led to larger estimation error and less stable coverage. Despite this deterioration, MV-FMR remained more reliable than separate U-FMR analyses.

\begin{table}[htbt]
\centering
\resizebox{\textwidth}{!}{%
\begin{tabular}{@{}c l cccc cccc@{}}
\toprule
& & \multicolumn{4}{c}{MV-FMR} & \multicolumn{4}{c}{U-FMR} \\
\cmidrule(lr){3-6}\cmidrule(lr){7-10}
\rotatebox{90}{Sc.} & & ISE & ISB & CO (\%) & SEr & ISE & ISB & CO (\%) & SEr \\
\midrule
\multirow{2}{*}{1} 
& $\beta_1(t)$ & 0.009 (0.011) & 0.002 (0.001) & 85.4 (16.4) & 0.96 (0.08) & 0.185 (0.337) & 0.049 (0.043) & 71.2 (26.3) & 0.63 (0.06) \\
& $\beta_2(t)$ & 0.008 (0.009) & 0.002 (0.001) & 86.5 (15.2) & 1.07 (0.07) & 0.135 (0.266) & 0.028 (0.027) & 73.6 (25.8) & 0.64 (0.06) \\
\midrule
\multirow{2}{*}{2} 
& $\beta_1(t)$ & 0.006 (0.007) & 0.002 (0.001) & 75.9 (18.3) & 0.97 (0.09) & 0.002 (0.001) & 0.002 (0.001) & 24.0 (12.8) & 0.67 (0.20) \\
& $\beta_2(t)$ & 0.004 (0.004) & 0.001 (0.000) & 96.1 (13.3) & 1.12 (0.12) & 0.150 (0.275) & 0.034 (0.030) & 74.4 (27.4) & 0.64 (0.07) \\
\midrule
\multirow{2}{*}{3} 
& $\beta_1(t)$ & 0.005 (0.016) & 0.002 (0.001) & 57.9 (25.9) & 0.88 (0.28) & 0.224 (0.395) & 0.078 (0.072) & 48.7 (25.7) & 0.58 (0.04) \\
& $\beta_2(t)$ & 0.003 (0.003) & 0.001 (0.001) & 62.6 (25.2) & 1.07 (0.15) & 0.121 (0.395) & 0.020 (0.022) & 63.3 (26.1) & 0.48 (0.10) \\
\bottomrule
\end{tabular}%
}
\vspace{0.5em}
\caption{Comparison of MV-FMR and U-FMR performance across simulation Scenarios 1--3 for the continuous outcome. ISE: Integrated Squared Error (mean (SD) across simulation replications); ISB: Integrated Squared Bias (mean (SD) across time points); CO: mean pointwise 95\% bootstrap CI coverage rate, \%; SEr: SE calibration ratio (mean (SD) across time points).}
\label{tab:comparison_continuous}
\end{table}

\begin{figure}[h]
\centering 
\includegraphics[width=0.8\textwidth]{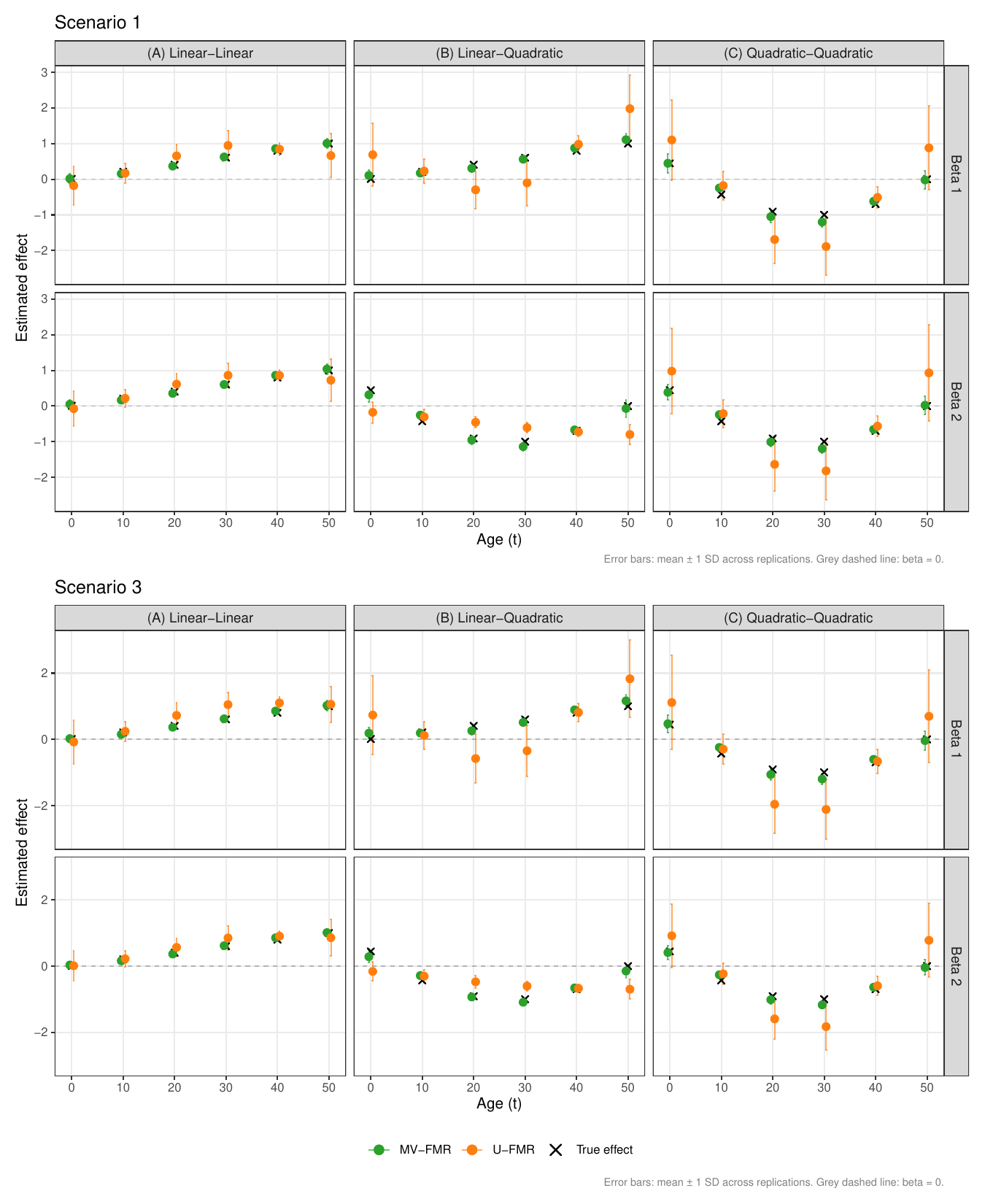}
\caption{Mean time-varying causal effect estimates at six representative time points, averaged across 500 simulation replications for a continuous outcome. Error bars show $\pm 1$ empirical standard deviation across replications.  Columns represent different exposure-outcome model combinations: (A)~linear--linear, (B)~linear--quadratic, (C)~quadratic--quadratic. Rows show the estimated values for $\beta_1(t)$ and $\beta_2(t)$. Estimates are colour-coded by method: green circles for MV-FMR and orange circles for U-FMR (benchmark); black crosses indicate the true causal effects. MV-FMR and U-FMR estimates are slightly offset horizontally for visual clarity. Top panel: Scenario~1 (pleiotropy). Bottom panel: Scenario~3 (mediation).}
\label{fig:sim_combined_beta}
\end{figure}

We then evaluated the robustness of the MV-FMR model to different functional forms of the exposure-outcome relationships for Scenarios 1 and 3. Specifically, we considered three combinations: (i) linear effects for both exposures, (ii) a linear effect for one exposure and a quadratic effect for the other, and (iii) quadratic effects for both exposures. To complement the summary metrics reported in Table~\ref{tab:comparison_continuous}, Figure~\ref{fig:sim_combined_beta} presents the mean estimated coefficient functions $\beta_1(t)$ and $\beta_2(t)$ at six representative time points, averaged across all the simulation replications for a continuous outcome. U-FMR estimates are included as a benchmark to illustrate the pointwise behaviour of the two approaches over time. Corresponding results for a binary outcome are provided in Supplementary Figure~1. Across the time points considered, the MV-FMR model recovered the true functional effects more accurately than the univariable approach. In particular, it captured nonlinear time-varying effects, whereas the univariable approach was less able to reproduce the underlying functional relationships. Supplementary Figures~2 and~3 show the pointwise ISE and ISB profiles over time for the three scenarios under linear true effects. Additional sensitivity analyses considered stepwise effects and decreasing lifetime effects, with results reported in Supplementary Table~3. MV-FMR showed stable performance across these additional specifications, supporting the robustness of the method across a wider range of true effect shapes.

We further investigated the performance of the MV-FMR model under different degrees of instrument overlap, mediation strength, and longitudinal data sparsity. Results for the continuous outcome are presented in Table~\ref{tab:sensitivity_ise_continuous}, with corresponding coverage results reported in Supplementary Table~4. Results for the binary outcome are reported in Supplementary Table~5. In Scenario~1, the proportion of overlapping instruments, $p_{12}$, was varied across 0.15, 0.30, 0.50, and 0.80 to evaluate performance as the sharing of genetic instruments across exposures increased. The mean ISE for both $\beta_1(t)$ and $\beta_2(t)$ remained low across all levels of instrument overlap. In Scenario~3, we assessed sensitivity to mediation strength by varying $\gamma$ across 0.15, 0.30, 0.50, and 0.80, where $\gamma$ controls the strength of the causal pathway from $X_1(t)$ to $X_2(t)$. MV-FMR recovered the direct causal effects with comparable ISE across the mediation strengths considered, suggesting that the method can separate direct from mediated effects in the scenarios considered. 
To evaluate the impact of longitudinal data sparsity, we varied the number of observation points per individual exposure across $\{5,10,25,50\}$ over the interval $[0,50]$ under Scenario~1. Estimation accuracy improved as the number of longitudinal measurements increased, with lower ISE for both $\beta_1(t)$ and $\beta_2(t)$ under denser sampling. Overall, these sensitivity analyses suggest that MV-FMR model remained stable under increasing instrument overlap and mediation strength, while denser longitudinal sampling improved recovery of the functional effect shapes.

\begin{table}[h]
\centering
\small
\setlength{\tabcolsep}{7pt}
\renewcommand{\arraystretch}{1.18}
\begin{tabular}{@{}lcccc@{}}
\toprule

\multicolumn{5}{@{}l}{A. Overlapping instruments, $p_{12}$, Scenario 1} \\
\addlinespace[0.15em]
$p_{12}$ 
& \textbf{0.15} & 0.30 & 0.50 & 0.80 \\
\cmidrule(lr){1-5}
$\beta_1(t)$ 
& 0.009 (0.011) & 0.011 (0.012) & 0.012 (0.015) & 0.016 (0.017) \\
$\beta_2(t)$ 
& 0.008 (0.009) & 0.010 (0.011) & 0.011 (0.013) & 0.017 (0.024) \\

\addlinespace[0.25em]
\midrule
\addlinespace[0.15em]

\multicolumn{5}{@{}l}{B. Mediation strength, $\gamma$, Scenario 3} \\
\addlinespace[0.15em]
$\gamma$ 
& 0.15 & \textbf{0.30} & 0.50 & 0.80 \\
\cmidrule(lr){1-5}
$\beta_1(t)$ 
& 0.004 (0.013) & 0.005 (0.016) & 0.006 (0.022) & 0.006 (0.007) \\
$\beta_2(t)$ 
& 0.004 (0.007) & 0.003 (0.003) & 0.003 (0.003) & 0.002 (0.002) \\

\addlinespace[0.25em]
\midrule
\addlinespace[0.15em]

\multicolumn{5}{@{}l}{C. Longitudinal sparsity, $n_{\text{sparse}}$, Scenario 1} \\
\addlinespace[0.15em]
$n_{\text{sparse}}$ 
& 5 & \textbf{10} & 25 & 50 \\
\cmidrule(lr){1-5}
$\beta_1(t)$ 
& 0.069 (0.102) & 0.009 (0.011) & 0.002 (0.002) & 0.001 (0.001) \\
$\beta_2(t)$ 
& 0.071 (0.138) & 0.008 (0.009) & 0.002 (0.001) & 0.001 (0.001) \\
\bottomrule
\end{tabular}
\vspace{0.5em}
\caption{Sensitivity of MV-FMR under different degrees of overlapping instruments, mediation strength, and longitudinal sparsity for the continuous outcome. Values are mean Integrated Squared Error, ISE, reported as mean (SD) across simulation replications, for $\beta_1(t)$ and $\beta_2(t)$. Bold values indicate the reference setting used in the main simulation scenarios. Corresponding coverage results are reported in Supplementary Table~4.}
\label{tab:sensitivity_ise_continuous}
\end{table}

We performed a sensitivity analysis to evaluate the effect of the total number of genetic instruments on MV-FMR performance and on the corresponding time-averaged heritability. The number of instruments was varied across $P \in \{25,50,75,100\}$ under Scenario~1 with a continuous outcome and results are reported in Supplementary Table~6. MV-FMR showed stable performance across the range of instrument counts considered, corresponding to different controlled levels of heritability.

Finally, we performed a sensitivity analysis to assess how the performance of the MV-FMR model changes as the number of exposures increases. The number of exposures was varied across $m \in \{3,4,5\}$ under Scenario~1 with a continuous outcome. Results are reported in Supplementary Table~7. The model remained able to recover the true time-varying causal effects as the number of exposures increased, supporting its applicability beyond the bivariable setting in the scenarios considered.

\section{Case study: Systolic Blood Pressure and Body Mass Index as time-varying exposure on Coronary Artery Disease}\label{sec:application}
We applied the proposed MV-FMR framework to investigate the time-varying causal effects of systolic blood pressure (SBP) and body mass index (BMI) on coronary artery disease (CAD) risk.
This application provides a good case study for our functional extension of multivariable MR for several reasons. First, SBP and BMI are well-established cardiovascular risk factors with substantial phenotypic correlation~\cite{hall2015obesity, collaboration2016body}, creating a scenario in which multivariable approaches offer advantages over univariable methods. Second, biological evidence indicates that BMI influences cardiovascular outcomes both directly and indirectly through blood pressure~\cite{seravalle2014obesity}, establishing a mediation pathway that traditional univariable MR cannot disentangle. Indeed, cross-sectional multivariable MR studies have shown that approximately 27\% of the BMI-CAD association operates through SBP~\cite{Gill2021}, providing a well-characterized benchmark for validation. Third, both exposures exhibit time-varying effects over the life course, with evidence suggesting that critical exposure windows for cardiovascular risk may differ between midlife and later life~\cite{Gnatiuc2017,duncan2021differential}. By jointly modelling SBP and BMI trajectories while accounting for their correlation and potential mediation, MV-FMR enables estimation of the direct time-varying causal effect of each exposure on CAD risk, allowing us to recover established causal patterns and uncover critical windows of effect that cannot be identified using standard cross-sectional or univariable approaches.

To address this question, we used data from UK Biobank, a large-scale prospective cohort study that recruited over 500,000 participants aged 40-69 years between 2006 and 2010 across the United Kingdom~\cite{Sudlow2015}. For this analysis, we restricted to participants who met the following criteria: (i) European ancestry, (ii) availability of genetic data, and (iii) at least two longitudinal measurements of both SBP and BMI between ages 50 and 70 years and before the CAD incidence. The 50-70 year age window was chosen to ensure sufficient longitudinal coverage of both SBP and BMI. Although UK Biobank recruited participants from age 40 onwards, measurements during the first decade were relatively sparse, reflecting the fact that laboratory values were mainly obtained from linked primary care records and were not uniformly available across the earlier adult age range. The temporal distribution of SBP and BMI measurements across ages 40-80 years among UK Biobank participants with European ancestry and genetic data is shown in Supplementary Figure~4. After applying these criteria, 59,907 participants were included in the final sample. SBP and BMI trajectories were reconstructed from all available measurements in primary care records and UK Biobank baseline assessment visits within the 50-70 year window. Among these participants, the median number of measurements per individual within the 50-70 year window was 4 (IQR: 2-16) for SBP and 4 (IQR: 2-8) for BMI. Although individual-level trajectories were sparse and irregularly spaced, pooling measurements across participants provided dense temporal coverage over the analysis window, as shown in Supplementary Figure~5. Incident CAD was defined using first-occurrence ICD-10 codes I21-I25, covering acute myocardial infarction and chronic ischemic heart disease.

\begin{figure}[h] 
    \centering
    \includegraphics[width=0.6\textwidth]{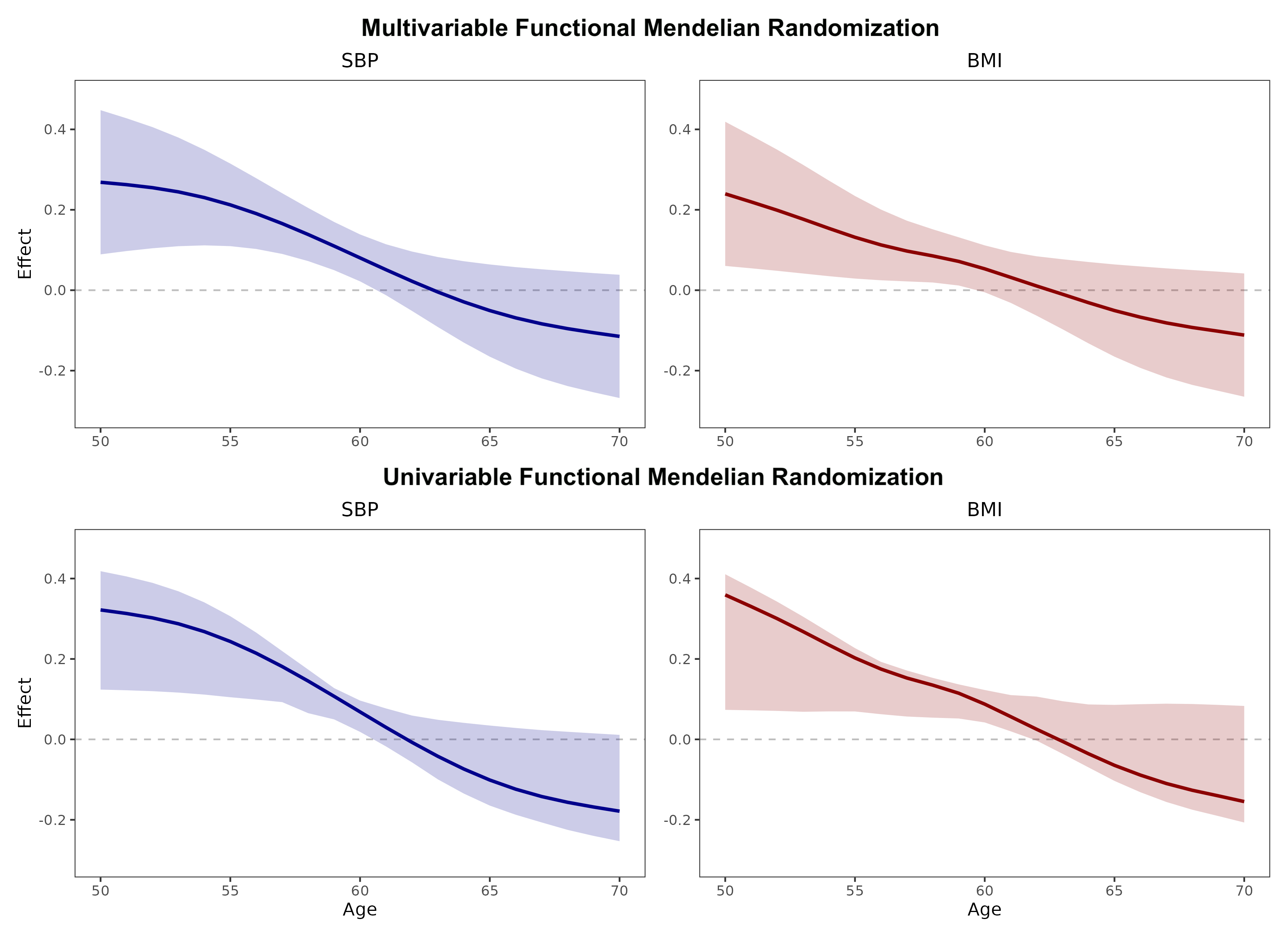} 
    \caption{Estimated time-varying causal effects of SBP and BMI on CAD risk from MV-FMR (top) and U-FMR (bottom). Blue curves show SBP effects with 95\% pointwise bootstrap CI; red curves show BMI effects with 95\% pointwise bootstrap CI.}
    \label{fig:application} 
\end{figure}

We investigated the effects of SBP and BMI on CAD using 274 and 199 genetic variants, respectively. Genetic instruments were obtained from large-scale genome-wide association study (GWAS) meta-analyses: those for SBP were taken from Evangelou et al.~\cite{Evangelou2018}, and those for BMI from Winkler et al.~\cite{Winkler2015}. Five variants were associated with both exposures, representing overlapping instruments that may introduce pleiotropy between SBP and BMI. We then applied MV-FMR to jointly estimate the time-varying causal effects of SBP and BMI on CAD risk over ages 50-70 years. Individual exposure trajectories were modelled using FPCA with PACE, with the number of components chosen via our data-driven k-fold cross-validation procedure. Two principal components were selected for both exposures. For SBP, the first two components explained 77.0\% and 16.5\% of the variability, respectively, while for BMI, they explained 90.5\% and 8.1\%. The large proportion of explained variance indicates that trajectories of both exposures were well captured by the two-component FPCA representation. Instrument strength was assessed using conditional F-statistics for each principal component. For SBP, the first and second components yielded conditional F-statistics of 9.77 and 3.44, respectively, while for BMI, the corresponding values were 9.03 and 2.57. These values indicate limited conditional instrument strength, particularly for the second principal components. Consequently, the estimates may be affected by weak-instrument bias and should be interpreted cautiously. We therefore regard this empirical analysis primarily as a proof-of-concept under potentially weak identification, illustrating the application of MV-FMR to longitudinal epidemiological data rather than providing evidence of strong causal identification.
To assess the benefits of joint modelling, we compared MV-FMR with U-FMR applied separately to each exposure. This comparison allowed us to evaluate whether simultaneous modelling of SBP and BMI revealed different causal effect patterns compared to the univariable analyses.
Figure~\ref{fig:application} presents the MV-FMR estimates. The analysis showed distinct time-varying causal effects of SBP and BMI on CAD risk. Both exposures showed positive associations with CAD risk between ages 50 and 60, but their effects attenuated thereafter, becoming non-significant (95\% bootstrap CI) beyond age 60. These findings suggest that the critical exposure window for SBP and BMI in relation to CAD risk is concentrated in the sixth decade of life. However, given the limited conditional instrument strength and the pointwise nature of inference, this pattern should be interpreted cautiously and should not be considered evidence of a distinct causal exposure window.

The comparison between MV-FMR and U-FMR provides evidence consistent with established mediation patterns. For SBP, the multivariable and univariable analyses yielded broadly consistent results, both identifying similar effect magnitudes and time windows of significance. For BMI, MV-FMR estimates showed attenuation in the point estimates compared to the univariable approach, suggesting that part of BMI's apparent effect on CAD operates indirectly through blood pressure pathways. Inspection of the confidence intervals reveals partial overlap between MV-FMR and U-FMR estimates for BMI. Notably, this overlap is also observed in the traditional MV-MR analysis~\cite{Gill2021}, where the univariable BMI effect (OR 1.49, 95\% CI 1.39 to 1.60) and the multivariable effect (OR 1.34, 95\% CI 1.24 to 1.45) showed overlapping confidence intervals.

\section{Discussion}\label{sec:discussion}
In this work, we introduced MV-FMR, a multivariable functional Mendelian randomization framework for estimating time-varying causal effects of multiple longitudinal exposures on continuous and binary outcomes. The method brings together two strands of recent MR methodology: multivariable MR, which allows the direct effects of related exposures to be estimated jointly, and functional MR, which treats exposure histories as trajectories rather than as measurements at isolated time points. This combination is motivated by life-course epidemiological settings in which risk factors evolve together, may share genetic predictors, and may influence disease risk through age-dependent mechanisms. MV-FMR represents each longitudinal exposure using functional principal component analysis and then incorporates the resulting scores into a multivariable MR model. This provides a data-driven dimension-reduction strategy for sparse and irregular exposure measurements while preserving the continuous-time structure of the underlying trajectories. By modelling multiple exposure processes jointly, MV-FMR can estimate age-varying direct effects conditional on the other included exposures, rather than marginal effects of each exposure considered in isolation. Importantly, the target of inference is the exposure--outcome effect function. This distinguishes MV-FMR from related approaches in the longitudinal and heterogeneous-effects MR literature. Previous functional MR approaches have focused primarily on single exposures~\cite{cao2016timevarying, Tian2024}, whereas multivariable time-varying MR approaches have typically represented time using a limited number of discrete exposure measurements~\cite{sanderson2022timevarying}. More broadly, approaches such as context-stratified MR investigate whether MR estimates differ across external settings or populations; MV-FMR instead uses within-individual exposure histories to study how causal effects vary over the life course. The framework, therefore, provides a way to ask not only whether an exposure has a causal effect, but when in the life course that effect may be most relevant and whether it persists after accounting for other longitudinal exposures.

Extensive simulation studies evaluated the finite-sample performance of MV-FMR across a range of scenarios, including nonlinear effects, horizontal pleiotropy, and mediation. Under the configurations considered, the method generally recovered the underlying functional effect patterns and outperformed the corresponding univariable functional MR analyses. The use of a GMM-based estimation procedure provides a flexible framework for incorporating genetic instruments into the functional model, but it does not eliminate weak-instrument bias. Instrument strength should therefore be assessed using conditional F-statistics for the retained FPCA components, and results should be interpreted cautiously when conditional instrument strength is limited. Sensitivity analyses varying instrument overlap, mediation strength, exposure-outcome functional forms, data sparsity, the number of genetic instruments and the number of exposures demonstrated good performance of the method under diverse conditions. The practical relevance of the proposed MV-FMR framework was demonstrated through an application to UK Biobank data, examining the time-varying causal effects of SBP and BMI on CAD risk. This analysis was intended primarily as a proof-of-concept under potentially weak identification, illustrating how the proposed framework can reproduce patterns consistent with established causal structures while extending them to a functional setting. The results highlighted ages 50-60 as a period during which both SBP and BMI exerted their strongest estimated effects, with attenuated effects at older ages. The comparison between multivariable and univariable functional MR provides evidence consistent with established mediation mechanisms. While SBP estimates were largely unchanged between univariable and multivariable models, BMI effects were attenuated in the joint analysis, consistent with part of BMI’s association with CAD operating through blood pressure pathways. This pattern mirrors findings from traditional cross-sectional MV-MR model, where partial mediation of the BMI-CAD relationship by SBP has been reported and where univariable and multivariable estimates similarly exhibit overlapping confidence intervals~\cite{Gill2021}. Taken together, these findings are consistent with the expected behaviour of MV-FMR in the presence of mediation, while providing a functional representation of effects across age.

Several limitations should be acknowledged. First, the MV-FMR model requires longitudinal measurements for each exposure over a common age range, which may not be available in all datasets. In addition, the computational burden increases with the number of exposures and retained FPCA components, because the method effectively expands each longitudinal exposure into multiple pseudo-exposures. Second, although some FPCA methods, such as PACE, can accommodate sparse and irregular longitudinal measurements, reliable estimation of the mean function, covariance surface, and eigenfunctions still depends on sufficient coverage of the time domain after pooling observations across individuals. To assess the impact of sparsity, we conducted dedicated simulation analyses varying the number of longitudinal observations per individual. These simulations showed that MV-FMR retained adequate estimation performance across the sparsity settings considered, including scenarios with only five observations per individual, supporting its applicability to sparse longitudinal data when the pooled measurement process provides sufficient temporal coverage. Third, uncertainty quantification may be affected by the data-driven selection of the number of retained FPCA components. Since the number of retained components is selected by cross-validation to optimize predictive performance, truncation bias may still be relevant for inference if it is of comparable magnitude to the standard error. Standard bootstrap procedures mainly capture sampling variability and may therefore not fully account for this bias, potentially contributing to the variation in pointwise coverage observed across scenarios and outcome types~\citep{hardle1988bootstrapping,hall2013simple,calonico2018effect}. Fourth, weak instruments can bias MR estimates toward the confounded observational association, particularly in finite samples. This issue is especially relevant for MV-FMR, because instrument strength must be assessed for the FPCA scores conditionally on the other functional components and exposures included in the model. Higher-order FPCA components often explain less exposure variation and may therefore be more weakly instrumented; as the number of retained components or functional exposures increases, conditional F-statistics may decline~\citep{sanderson2016weak}. Weak-instrument bias may also contribute to undercoverage, particularly when conditional instrument strength is limited. Although MV-FMR performed adequately under the instrument configurations considered in our simulations, these results should not be interpreted as demonstrating general robustness to weak instruments. Reporting conditional instrument strength and assessing uncertainty calibration in application-specific simulations are, therefore, important before interpreting MV-FMR estimates. Fifth, the one-sample implementation of MV-FMR may be affected by sample overlap bias, particularly in the presence of weak instruments. The two-sample extension described in Supplementary Method~3 can address this source of bias when independent SNP-outcome associations are available. However, unlike standard two-sample MR, MV-FMR also requires SNP associations with the FPCA scores of the functional exposures, which need to be estimated from individual-level longitudinal exposure data and are not currently available in most open GWAS repositories. Bias-correction methods developed for MV-MR model, such as MRBEE, may provide a useful starting point for future work on weak-instrument and sample-overlap bias correction within the functional data analysis framework~\citep{lorincz2024mrbee}. Sixth, MV-FMR remains vulnerable to residual horizontal pleiotropy, whereby genetic variants affect the outcome through pathways other than the modelled functional exposures. Modelling multiple exposures can account for pleiotropic pathways operating through the included risk factors, but it cannot rule out residual pleiotropy through unmeasured traits. Overidentification and heterogeneity diagnostics may reveal incompatibility with the assumed moment conditions, but violations of the exclusion restriction are not directly testable. Future extensions incorporating pleiotropy-robust approaches, such as MR-Egger, within the functional data analysis framework may help address this limitation~\citep{bowden2015mendelian}. 
Seventh, inference in the current implementation is pointwise, and we do not provide a formal global test of whether the causal effect function is null over the entire time domain, nor a simultaneous confidence band for the effect function. Consequently, Scenario 2 assesses pointwise behaviour under a null effect function rather than global type-I error control. Developing global inference for MV-FMR, potentially by adapting variance-component score-test approaches for nonparametric functions~\citep{liu2007semiparametric}, represents an important direction for future work. Finally, although the empirical application demonstrates the feasibility of MV-FMR, it was primarily intended as a proof of concept. The analysis was restricted to ages 50-70 and conditioned on survival to age 50 without prevalent CAD, which may select a cardiovascularly resilient subset of the population. Individuals at higher risk may develop CAD earlier and therefore be excluded from the baseline risk set. While MR mitigates traditional confounding, it does not eliminate survival or selection bias, which may attenuate estimated effects at older ages. Future applications incorporating detailed prescription records, broader age ranges, and sensitivity analyses considering the timing of outcome measurement relative to exposure trajectories will be important to assess the practical utility and robustness of MV-FMR across epidemiological settings.

In summary, MV-FMR represents a methodological advance in functional Mendelian Randomization, offering a flexible and interpretable framework for estimating time-varying causal effects in multivariable settings. By integrating functional data analysis with multivariable MR, the method enables the identification of life-course critical periods, the disentangling of mediation pathways, and an improved understanding of the dynamic interplay between correlated risk factors. Future work should investigate computational strategies for large-scale biobank applications. MV-FMR holds promise for generating richer causal insights and informing more precise preventive interventions in population health.

\paragraph{Data sharing}
UKB data are accessible for health-related research in the public interest. This study used data from the UKB Resource under Application Number 102297.

\paragraph{Acknowledgements}
No funding was received for this research.\\
The present research is part of the activities of “Dipartimento di Eccellenza 2023-2027". 
F. Ieva acknowledges the National Plan for NRRP Complementary Investments "Advanced Technologies for Human-centred Medicine" (PNC0000003).

\textit{A large language model (Claude) was used solely to improve language, grammar, and readability of selected portions of the manuscript. All scientific content, methodological development, analyses, and interpretations were produced entirely by the authors.}

\bibliographystyle{unsrt}   
\bibliography{references}   

\clearpage
\raggedbottom
\setcounter{equation}{0}
\setcounter{figure}{0}
\setcounter{table}{0}
\renewcommand{\theequation}{S\arabic{equation}}
\renewcommand{\thefigure}{S\arabic{figure}}
\renewcommand{\thetable}{S\arabic{table}}
\renewcommand{\theHequation}{supp.\arabic{equation}}
\renewcommand{\theHfigure}{supp.\thefigure}
\renewcommand{\theHtable}{supp.\thetable}
\pdfbookmark[0]{Supplementary Material}{supplementary}
\begin{center}
{\Large\bfseries Supplementary Material\par}
\vspace{1em}
{\large\bfseries Unraveling time-varying causal effects of multiple exposures: integrating Functional Data Analysis with Multivariable Mendelian Randomization\par}
\vspace{1em}
Nicole Fontana, Francesca Ieva, Luisa Zuccolo, Emanuele Di Angelantonio, Piercesare Secchi
\end{center}
\vspace{1em}

\textbf{Supplementary Methods}

\textbf{Supplementary Method 1:} Functional Principal Component Analysis with PACE method \vspace{0.5em}

\textbf{Supplementary Method 2:} Bootstrap inference
\vspace{0.5em}

{\color{black}\textbf{Supplementary Method 3:} Two-sample extension of MV-FMR}
\vspace{0.5em}

{\color{black}\textbf{Supplementary Method 4:} Heritability of simulated functional exposures} \vspace{0.5em}

\vspace{1em}
\textbf{Supplementary Tables} 

\textbf{Supplementary Table 1:} Conditional F-statistics across simulation scenarios 1–3 \vspace{0.5em}

{\color{black}\textbf{Supplementary Table 2:} ISE, ISB, coverage rate, and SE calibration of MV-FMR and U-FMR across simulation scenarios 1–3 for binary outcomes} \vspace{0.5em}

{\color{black}\textbf{Supplementary Table 3:} Sensitivity analysis of MV-FMR performance under alternative exposure-outcome models} \vspace{0.5em}

\textbf{Supplementary Table 4:} Sensitivity analysis of MV-FMR performance (coverage) under alternative overlapping instruments, mediation strength, and longitudinal sparsity for the continuous outcome \vspace{0.5em}

\textbf{Supplementary Table 5:} Sensitivity analysis of MV-FMR performance under alternative overlapping instruments, mediation strength, and longitudinal sparsity for the binary outcome \vspace{0.5em}

{\color{black}\textbf{Supplementary Table 6:} Sensitivity analysis of MV-FMR performance under varying number of genetic instruments} \vspace{0.5em}

{\color{black}\textbf{Supplementary Table 7:} Sensitivity analysis of MV-FMR performance under varying number of exposures} \vspace{0.5em}

\vspace{1em}
\textbf{Supplementary Figures} 

\textbf{Supplementary Figure 1:} Mean time-varying causal effect estimates across different exposure-outcome model combinations with a binary outcome \vspace{0.5em}

{\color{black}\textbf{Supplementary Figure 2:} Mean pointwise integrated squared error at six representative time points (continuous outcome)} \vspace{0.5em}

{\color{black}\textbf{Supplementary Figure 3:} Pointwise integrated squared bias at six representative time points (continuous outcome)} \vspace{0.5em}

{\color{black}\textbf{Supplementary Figure 4:} Distribution of pooled SBP and BMI measurement ages among UK Biobank participants (ages 40–80)} \vspace{0.5em}

{\color{black}\textbf{Supplementary Figure 5:} Distribution of pooled SBP and BMI measurement ages within the analysis cohort (ages 50–70)} \vspace{0.5em}

\newpage
\section*{Supplementary Method 1: Functional Principal Component Analysis with PACE method}
Functional Principal Component Analysis (FPCA) provides a low-dimensional representation of smooth functional data. For each exposure $X_j(t)$ observed over $t \in [0,T]$, let $\mu_j(t)$ denote the mean trajectory and $\text{cov}(X_j(s), X_j(t))$ the covariance function. The eigendecomposition of the covariance operator yields:
\begin{equation}
\text{cov}(X_j(s), X_j(t)) = \sum_{k=1}^{\infty} \lambda_{j,k} \phi_{j,k}(s) \phi_{j,k}(t),
\end{equation}
where $\lambda_{j,1} \ge \lambda_{j,2} \ge \dots \ge 0$ are the eigenvalues and $\phi_{j,k}(t)$ are the orthonormal eigenfunctions. By the Karhunen-Loève theorem, each individual trajectory can be written as:
\begin{equation}\label{supp:eq:kl_expansion}
X_j(t) = \mu_j(t) + \sum_{k=1}^{\infty} \xi_{j,k} \phi_{j,k}(t),
\end{equation}
where $\xi_{j,k} = \int_0^T (X_j(t) - \mu_j(t)) \phi_{j,k}(t) dt$ are uncorrelated scores with variance $var(\xi_{j,k}) = \lambda_{j,k}$ and $cov(\xi_{j,k_1},\xi_{j,k_2})= 0$ for $k_1 \neq k_2$.

In practice, longitudinal data are often sparse and irregularly sampled. To address this, we implement the Principal Analysis by Conditional Estimation (PACE) method~\cite{Yao2005}, modelling observed measurements as $X_{j,i}(t_{j,i,l}) = X^*_j(t_{j,i,l}) + \epsilon_{j,i,l}$, where $X^*_j(t)$ is the true underlying trajectory and $\epsilon_{j,i,l} \stackrel{\text{iid}}{\sim} \mathcal{N}(0, \sigma_j^2)$ represents measurement error. Under the assumption that individual trajectories follow the FPCA representation above, the principal component scores are estimated via conditional expectation.
\begin{equation}\label{supp:eq:pace_scores}
\hat{\xi}_{j,k} = \mathbb{E}[\xi_{j,k} | \mathbf{X}_{j}] = \lambda_{j,k} \boldsymbol{\phi}_{j,k}^T \boldsymbol{\Sigma}_{j}^{-1}(\mathbf{X}_{j} - \boldsymbol{\mu}_{j}).
\end{equation}
\textcolor{black}{The expansion in~\eqref{supp:eq:kl_expansion} is truncated at $K_j$ components for each exposure $j$, with $K_j$ selected via a data-driven, greedy forward-selection procedure evaluated by 5-fold cross-validation. Starting from $K_j = 2$ retained components for each exposure, candidate models are generated by increasing the number of retained components for one exposure at a time while keeping the others fixed. Each candidate model is refitted on the training folds and evaluated out of sample using mean squared error for continuous outcomes and area under the ROC curve for binary outcomes. The component counts are therefore selected jointly across exposures within a single cross-validation procedure, rather than separately from exposure-specific models. At each step, the increase yielding the largest relative improvement in the cross-validation criterion is retained if the improvement exceeds 0.1\%; otherwise, the procedure stops and the current values of $K_j$ are retained. This threshold was used consistently across the simulation results and the UK Biobank application. Candidate values are bounded by the number of FPCA components available for each exposure and by a combined maximum of $K_{\max}$, fixed to $6\cdot m$ in our case, retained components across exposures.}

\newpage
\section*{Supplementary Method 2: Bootstrap inference}
Statistical inference for the time-varying causal effect functions $\beta_j(t)$ is conducted using case resampling bootstrap with $B=100$ replications. For each bootstrap sample $b = 1, \ldots, B$:

\textbf{Step 1: Resample individuals with replacement}
\begin{itemize}
    \item Draw $N$ individuals with replacement from $\{1, \ldots, N\}$: 
    $\{i_1^{(b)}, \ldots, i_N^{(b)}\}$
    \item This creates bootstrap samples:
    $\{(Y_i^{(b)}, \boldsymbol{\xi}_i^{(b)}, \mathbf{G}_i^{(b)})\}$
    where $\boldsymbol{\xi}_i = [\xi_{i1,1}, \ldots, \xi_{i1,K_1},
    \ldots, \xi_{im,1}, \ldots, \xi_{im,K_m}]^T$ are the pre-estimated
    FPCA scores for all $m$ exposures
\end{itemize}

\textbf{Step 2: Re-estimate causal coefficients}
\begin{itemize}
    \item For continuous outcomes: estimate
    $\hat{\boldsymbol{\beta}}^{*(b)} = [\hat{\boldsymbol{\beta}}_1^{*(b)},
    \ldots, \hat{\boldsymbol{\beta}}_m^{*(b)}]$ via continuously updating GMM
    using bootstrap samples $(\boldsymbol{\xi}^{(b)}, Y^{(b)},
    \mathbf{G}^{(b)})$
    \item For binary outcomes apply 2SRI procedure: regress $\boldsymbol{\xi}^{(b)}$ on 
        $\mathbf{G}^{(b)}$ to obtain residuals $\hat{\mathbf{V}}^{(b)}$ and fit logistic regression with both 
        $\boldsymbol{\xi}^{(b)}$ and $\hat{\mathbf{V}}^{(b)}$ 
        to obtain $\hat{\boldsymbol{\beta}}^{*(b)}$
\end{itemize}

\textbf{Step 3: Reconstruct functional effects}
Using the eigenfunctions $\{\phi_{j,k}(t)\}$ from the original sample, reconstruct:
    $$
    \hat{\beta}_j^{(b)}(t) = \sum_{k=1}^{K_j} \hat{\beta}_{j,k}^{*(b)} 
    \phi_{j,k}(t).
    $$

After obtaining $B$ bootstrap replications of $\hat{\beta}_j(t)$ at each time point, pointwise confidence intervals are:
$$
\text{CI}_{0.95}(\beta_j(t)) = \left[\hat{\beta}_j^{(0.025)}(t), 
\hat{\beta}_j^{(0.975)}(t)\right],
$$
where $\hat{\beta}_j^{(q)}(t)$ is the $q$-th quantile of $\{\hat{\beta}_j^{(1)}(t), \ldots, \hat{\beta}_j^{(B)}(t)\}$.

\newpage
{\color{black}
\section*{Supplementary Method 3: Two-sample extension of MV-FMR}
The MV-FMR framework described in the main manuscript is formulated for a one-sample design, in which longitudinal exposure measurements and the outcome are available for the same individuals. As an extension, MV-FMR can also be implemented in a two-sample setting, where genetic associations with the functional exposure scores are estimated in an exposure sample and genetic associations with the outcome are obtained from an independent outcome sample. When the two samples are non-overlapping and drawn from comparable populations, this formulation avoids bias due to sample overlap and enables the use of large-scale GWAS summary statistics for the outcome. This section describes the corresponding summary-level formulation and the estimator implemented in the \textit{mvfmr} R package.

\paragraph{Inputs.}
Let $\mathcal{S}_X$ denote the exposure sample and $\mathcal{S}_Y$ the outcome sample. The two-sample implementation requires:
\begin{itemize}
    \item from $\mathcal{S}_X$: individual-level genotype data for the selected instruments, $\mathbf{G} \in \mathbb{R}^{N_X \times P}$, and the FPCA results for each exposure, including the estimated scores $\hat{\xi}_{j,k_j}$ and eigenfunctions $\hat{\phi}_{j,k_j}(t)$, obtained as described in Supplementary Method~1;
    \item from $\mathcal{S}_Y$: for each instrument $l = 1,\ldots,P$, the estimated SNP--outcome association $\hat{\gamma}_l$, its standard error $\hat{s}_l$, and the outcome GWAS sample size $N_Y$.
\end{itemize}
Here, $j=1,\ldots,m$ indexes the functional exposures, $k_j=1,\ldots,K_j$ indexes the retained FPCA components for exposure $j$, and $l=1,\ldots,P$ indexes the genetic instruments. The total number of FPCA-derived pseudo-exposures is denoted by $K=\sum_{j=1}^{m}K_j$.

\paragraph{SNP--FPCA score associations.}
For each exposure $j$, component $k_j$, and instrument $l$, the SNP-FPCA score association is estimated in $\mathcal{S}_X$ by regressing the estimated FPCA score $\hat{\xi}_{j,k_j}$ on the genotype $G_l$. This gives the estimated association $\hat{b}_{l,j,k_j}$. Stacking these associations over all retained components and exposures yields the matrix:

\[
\hat{\mathbf{B}}_X \in \mathbb{R}^{P \times K}.
\]
The corresponding vector of SNP--outcome associations from $\mathcal{S}_Y$ is denoted by:
\[
\hat{\boldsymbol{\gamma}} =
(\hat{\gamma}_1,\ldots,\hat{\gamma}_P)^T.
\]

\paragraph{Two-sample MV-FMR estimator.}
The causal effects of the FPCA-derived pseudo-exposures are estimated by an inverse-variance weighted summary-level estimator:
\[
\hat{\boldsymbol{\beta}}^{*}
=
\left(
\hat{\mathbf{B}}_X^T
\mathbf{W}
\hat{\mathbf{B}}_X
\right)^{-1}
\hat{\mathbf{B}}_X^T
\mathbf{W}
\hat{\boldsymbol{\gamma}},
\]
where:
\[
\mathbf{W}
=
\mathrm{diag}
\left(
\frac{1}{\hat{s}_1^2},\ldots,\frac{1}{\hat{s}_P^2}
\right)
\]
is the inverse-variance weighting matrix based on the outcome GWAS standard errors. This estimator is the summary-level analogue of the multivariable MR estimator applied to the stacked FPCA score matrix $\boldsymbol{\xi}=[\boldsymbol{\xi}_1,\ldots,\boldsymbol{\xi}_m]$.

Under the inverse-variance weighting scheme defined above, and treating $\hat{\mathbf{B}}_X$ as fixed, the variance of $\hat{\boldsymbol{\beta}}^{*}$ is estimated as:
\[
\widehat{\mathrm{Var}}(\hat{\boldsymbol{\beta}}^{*})
=
\left(
\hat{\mathbf{B}}_X^T
\mathbf{W}
\hat{\mathbf{B}}_X
\right)^{-1}.
\]
This variance estimate accounts for sampling uncertainty in the SNP-outcome associations only, and does not propagate uncertainty in the estimated SNP-FPCA score associations or in the FPCA step.

\paragraph{Reconstruction and pointwise inference.}
The estimated time-varying causal effect for exposure $j$ is reconstructed as
\[
\hat{\beta}_j(t)
=
\boldsymbol{\phi}_j^T(t)
\hat{\boldsymbol{\beta}}_j^{*},
\qquad j=1,\ldots,m,
\]
where:
\[
\boldsymbol{\phi}_j(t)
=
[
\hat{\phi}_{j,1}(t),\ldots,\hat{\phi}_{j,K_j}(t)
]^T
\]
and $\hat{\boldsymbol{\beta}}_j^{*}$ is the subvector of $\hat{\boldsymbol{\beta}}^{*}$ corresponding to exposure $j$. Pointwise standard errors are obtained as:
\[
\hat{\sigma}_j^2(t)
=
\boldsymbol{\phi}_j^T(t)
\widehat{\mathrm{Var}}(\hat{\boldsymbol{\beta}}_j^{*})
\boldsymbol{\phi}_j(t),
\]
where $\widehat{\mathrm{Var}}(\hat{\boldsymbol{\beta}}_j^{*})$ is the corresponding submatrix of $\widehat{\mathrm{Var}}(\hat{\boldsymbol{\beta}}^{*})$. Pointwise 95\% confidence intervals are then given by:
\[
\hat{\beta}_j(t) \pm 1.96\,\hat{\sigma}_j(t).
\]

\paragraph{Implementation.}
The two-sample estimator is implemented in the \textit{mvfmr} R package through:
\begin{itemize}
    \item \texttt{fmvmr\_twosample()} for joint multivariable two-sample MV-FMR;
    \item \texttt{fmvmr\_separate\_twosample()} for separate univariable two-sample FMR analyses for each exposure.
\end{itemize}
Both functions require individual-level genotype data and FPCA objects from $\mathcal{S}_X$, together with summary-level SNP--outcome associations, their standard errors, and the outcome GWAS sample size from $\mathcal{S}_Y$.}

{\color{black}
\newpage
\section*{Supplementary Method 4: Heritability of simulated functional exposures}

We provide an analytical characterization of the heritability of the simulated functional exposures, defined pointwise in time as the proportion of total variance in $X_j(t)$ attributable to genetic variation:
\begin{equation*}
    h^2_j(t) = \frac{\mathrm{Var}\!\left(\sum_{l=1}^{P} \alpha_{l}(t)\, G_l\right)}{\mathrm{Var}(X_j(t))}.
\end{equation*}

\noindent\textbf{Data-generating process.} The functional exposure $X_j(t)$ is generated as:
\begin{equation*}
    X_j(t) = \sum_{l=1}^{P} \alpha_{l}(t)\, G_{l} + \bigl[U_{0j} + U_j(t)\bigr] + \epsilon_{j}(t),
\end{equation*}
where $G_l \sim \mathrm{Binomial}(2,\, 0.3)$ denotes the $l$-th genetic variant; the time-varying genetic effect is $\alpha_{l}(t) = a_{l} + b_{l}\cdot t$ with $a_{l} \sim \mathrm{Uniform}(-0.1,\,0.1)$ and $b_{l} \sim \mathrm{Uniform}(-0.004,\,0.004)$; $U_{0j} \sim \mathcal{N}(0,1)$ is a time-fixed confounder and $U_j(t)$ is a time-varying confounder modelled as a discretized Wiener process with $\mathrm{Var}(U_j(t)) = t/T$; and $\epsilon_{j}(t)$ is exposure-specific residual noise, also a discretized Wiener process with $\mathrm{Var}(\epsilon_{j}(t)) = t/T$, where $T = 50$.

\noindent\textbf{Variance components.} Since the $G_l$ are independent and $\mathrm{E}[\alpha_{l}(t)] = 0$:
\begin{align*}
    \mathrm{Var}\!\left(\sum_{l=1}^{P} \alpha_{l}(t)\, G_l\right) 
    &= P \cdot \mathrm{E}\!\left[\alpha_{l}(t)^2\right] \cdot \mathrm{Var}(G_l) \\
    &= P \left(\frac{(0.1)^2}{3} + t^2\,\frac{(0.004)^2}{3}\right) \cdot (2 \times 0.3 \times 0.7).
\end{align*}
The total variance of $X_j(t)$ is:
\begin{equation*}
    \mathrm{Var}(X_j(t)) = P \cdot \mathrm{E}\!\left[\alpha_{l}(t)^2\right] \cdot \mathrm{Var}(G_l)
    + 1 + \frac{2t}{T}.
\end{equation*}

\noindent\textbf{Numerical evaluation.} Substituting $P = 50$, $\mathrm{Var}(G_l) = 0.42$, and $T = 50$, the pointwise heritability at three representative time points is:

\noindent\textit{At $t = 0$:}
\begin{align*}
    \mathrm{Var}_{\mathrm{gen}}(0) &= 50 \times
    \frac{0.01}{3} \times 0.42 \approx 0.070, \\
    \mathrm{Var}_{\mathrm{total}}(0) &\approx 0.070 + 1 + 0 = 1.070, \\
    h^2(0) &\approx \frac{0.070}{1.070} \approx 0.065.
\end{align*}

\noindent\textit{At $t = 25$:}
\begin{align*}
    \mathrm{Var}_{\mathrm{gen}}(25) &= 50  \times
    \left(\frac{0.01}{3} + 625 \times \frac{0.000016}{3}\right) \times 0.42
    \approx 0.140, \\
    \mathrm{Var}_{\mathrm{total}}(25) &\approx 0.140 + 1 + 1 = 2.140, \\
    h^2(25) &\approx \frac{0.140}{2.140} \approx 0.065.
\end{align*}

\noindent\textit{At $t = 50$:}
\begin{align*}
    \mathrm{Var}_{\mathrm{gen}}(50) &= 50  \times
    \left(\frac{0.01}{3} + 2500 \times \frac{0.000016}{3}\right) \times 0.42
    \approx 0.350, \\
    \mathrm{Var}_{\mathrm{total}}(50) &\approx 0.350 + 1 + 2 = 3.350, \\
    h^2(50) &\approx \frac{0.350}{3.350} \approx 0.104.
\end{align*}

Averaging $h^2(t)$ over $[0, T]$ yields a heritability of approximately 0.071. This estimate confirms that the simulated functional exposures have heritability consistent with realistic values reported in MR simulation studies.\\

\noindent\textbf{Pointwise heritability as a function of age.} Figure~\ref{supp:fig:h2_curve} displays $h^2(t)$ over the full simulation domain $t \in [0, 50]$. A key feature of the functional MR setting is that heritability is itself time-varying. The genetic variance $\mathrm{Var}_{\mathrm{gen}}(t)$ grows quadratically in $t$, while the confounder and noise variance grow only linearly ($2t/T$). This produces a non-monotone pattern: $h^2(t)$ first declines slightly from $t=0$ (as the linear confounder term initially dominates), reaches a minimum of approximately 0.055 near $t \approx 10$, and then rises steadily to 0.104 at $t = 50$. 

\renewcommand{\thefigure}{SM4}
\begin{figure}[H]
\centering
\includegraphics[width=0.6\textwidth]{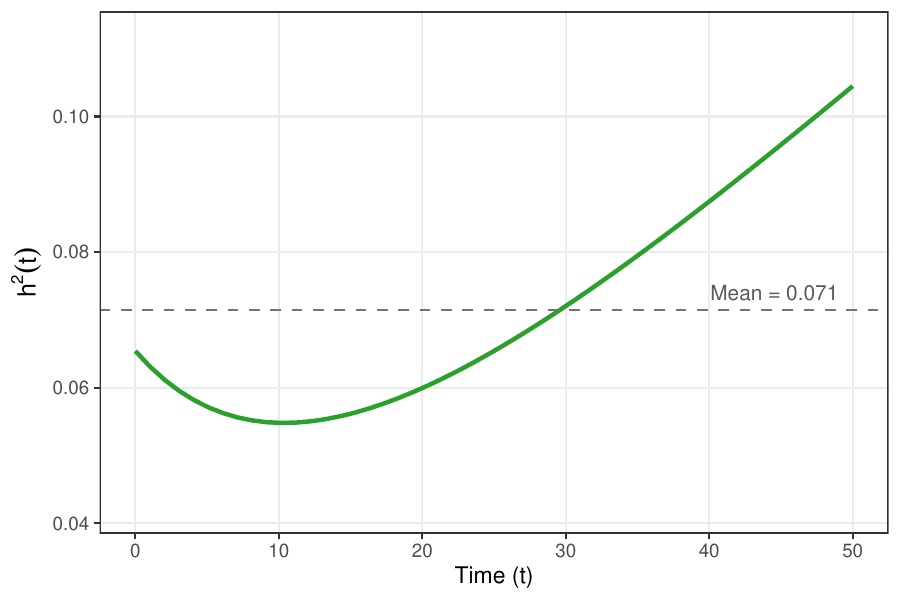}
\caption{Analytical pointwise heritability $h^2(t)$ of the simulated functional exposures over $t \in [0, 50]$. The curve is derived from the data-generating process defined above. Dashed grey line: time-averaged heritability ($\approx 0.071$).}
\label{supp:fig:h2_curve}
\end{figure}}
\renewcommand{\thefigure}{S\arabic{figure}}
\setcounter{figure}{0}

\newpage
\renewcommand{\thetable}{S1}
\setcounter{table}{0}
\section*{Supplementary Table 1: Conditional F-statistics across simulation scenarios 1–3}
\begin{table}[H]
\centering
\begin{tabular}{cccc|ccc}
\toprule
 & \multicolumn{3}{c|}{$X_1$ (cFF)} & \multicolumn{3}{c}{$X_2$ (cFF)} \\
\cmidrule(lr){2-4} \cmidrule(lr){5-7}
Scenario & PC1 & PC2 & PC3 & PC1 & PC2 & PC3 \\
\midrule
1 & 8.11 (1.48) & 6.99 (1.47) & 1.87 (0.37) & 8.06 (1.53) & 6.80 (1.43) & 1.87 (0.34) \\
2 & 8.11 (1.48) & 6.99 (1.47) & 1.87 (0.37) & 8.06 (1.53) & 6.80 (1.43) & 1.87 (0.34) \\
3 & 8.08 (1.31) & 7.10 (1.32) & 1.84 (0.34) & 5.37 (1.05) & 4.71 (0.93) & 1.46 (0.33) \\
\bottomrule
\end{tabular}
\caption{Conditional F-statistics (mean (SD) across replications) for each principal component (PC1--PC3) across the three simulation scenarios. Reported separately for $X_1$ and $X_2$.}
\label{supp:tab:fstats}
\end{table}

{\color{black}
\renewcommand{\thetable}{S2}
\setcounter{table}{0}
\section*{Supplementary Table 2: ISE, ISB, coverage rate, and SE calibration of MV-FMR and U-FMR across simulation scenarios 1–3 for binary outcomes}

\begin{table}[h!]
\centering
\resizebox{\textwidth}{!}{%
\begin{tabular}{@{}c l cccc cccc@{}}
\toprule
& & \multicolumn{4}{c}{MV-FMR} & \multicolumn{4}{c}{U-FMR} \\
\cmidrule(lr){3-6}\cmidrule(lr){7-10}
\rotatebox{90}{Sc.} & & ISE & ISB & CO (\%) & SEr & ISE & ISB & CO (\%) & SEr \\
\midrule
\multirow{2}{*}{1} 
& $\beta_1(t)$ & 0.029 (0.009) & 0.027 (0.022) & 32.6 (23.8) & 1.56 (0.20) & 0.239 (0.011) & 0.238 (0.209) & 2.4 (2.2) & 0.49 (0.04) \\
& $\beta_2(t)$ & 0.029 (0.009) & 0.028 (0.022) & 32.2 (24.2) & 1.78 (0.15) & 0.240 (0.009) & 0.240 (0.213) & 3.2 (3.6) & 0.58 (0.03) \\
\midrule
\multirow{2}{*}{2} 
& $\beta_1(t)$ & 0.007 (0.004) & 0.005 (0.003) & 57.7 (31.2) & 1.31 (0.22) & 0.012 (0.005) & 0.011 (0.007) & 23.5 (20.5) & 0.95 (0.07) \\
& $\beta_2(t)$ & 0.000 (0.001) & 0.000 (0.000) & 100.0 (0.0) & 2.40 (0.69) & 0.015 (0.056) & 0.001 (0.001) & 62.0 (36.2) & 0.26 (0.07) \\
\midrule
\multirow{2}{*}{3} 
& $\beta_1(t)$ & 0.042 (0.009) & 0.041 (0.035) & 23.8 (20.0) & 1.56 (0.32) & 0.216 (0.011) & 0.215 (0.188) & 2.7 (2.6) & 0.47 (0.03) \\
& $\beta_2(t)$ & 0.041 (0.009) & 0.039 (0.032) & 23.2 (19.4) & 1.66 (0.24) & 0.216 (0.012) & 0.215 (0.189) & 3.2 (3.4) & 0.43 (0.06) \\
\bottomrule
\end{tabular}%
}
\vspace{0.5em}
\caption{Comparison of MV-FMR and U-FMR performance across simulation Scenarios 1--3 for the binary outcome. ISE: Integrated Squared Error (mean (SD)); ISB: Integrated Squared Bias (mean (SD) across time points); CO: pointwise 95\% bootstrap CI coverage rate, \%; SEr: SE calibration ratio (mean (SD) across time points).}
\label{supp:tab:comparison_binary}
\end{table}
}

\renewcommand{\thetable}{S3.\arabic{table}}
\setcounter{table}{0}
\section*{Supplementary Table 3: Sensitivity analysis of MV-FMR performance under alternative exposure-outcome models}
Table~\ref{supp:tab:scen_def} summarizes the functional forms of the exposure-outcome relationships used in additional simulation analyses for Scenarios 1 and 3. Each row represents a specific combination of true time-varying effects $\beta_1(t)$ and $\beta_2(t)$, allowing evaluation of MV-FMR performance under alternative functional forms.

\begin{table}[H]
\centering
\small
\begin{tabular}{cl}
\toprule
Case & Functions $(\beta_1(t), \beta_2(t))$ \\
\midrule
A & $(0.4 - 0.02 \cdot t, \; 0.4 - 0.02\cdot t)$ \\
B & $(0.4 - 0.02 \cdot t, \; 0.1 \cdot \mathbf{1}(t<20))$ \\
C & $(0.1 \cdot \mathbf{1}(t<20), \; 0.4 - 0.02\cdot t)$ \\
D & $(0.1 \cdot \mathbf{1}(t<20), \; 0.1 \cdot \mathbf{1}(t<20))$ \\
\bottomrule
\end{tabular}
\caption{Functional forms of the exposure-outcome relationships used in additional simulation analyses for Scenarios 1 and 3. Each case specifies the true time-varying effects $\beta_1(t)$ and $\beta_2(t)$ considered to evaluate the robustness of MV-FMR under alternative functional forms.}
\label{supp:tab:scen_def}
\end{table}

\begin{table}[H]
\centering
\small
\setlength{\tabcolsep}{4pt}
\renewcommand{\arraystretch}{1.2}
\resizebox{\textwidth}{!}{%
\begin{tabular}{@{}c l l cccc cccc@{}}
\toprule
 & & & \multicolumn{4}{c}{$\beta_1(t)$} & \multicolumn{4}{c}{$\beta_2(t)$} \\
\cmidrule(lr){4-7}\cmidrule(lr){8-11}
\rotatebox{90}{Sc.} & Case & Method & ISE & ISB & CO (\%) & SEr & ISE & ISB & CO (\%) & SEr \\
\midrule
\multirow{8}{*}{1} & \multirow{2}{*}{A} & MV-FMR & 0.006 (0.005) & 0.003 (0.002) & 60.7 (13.5) & 0.81 (0.06) & 0.006 (0.003) & 0.003 (0.002) & 61.2 (10.8) & 0.92 (0.04) \\
 &  & U-FMR & 0.018 (0.023) & 0.006 (0.004) & 50.8 (31.0) & 0.62 (0.07) & 0.017 (0.025) & 0.005 (0.004) & 58.1 (29.7) & 0.69 (0.05) \\
 & \multirow{2}{*}{B} & MV-FMR & 0.005 (0.003) & 0.003 (0.003) & 51.1 (9.6) & 0.81 (0.07) & 0.002 (0.002) & 0.000 (0.001) & 88.2 (16.0) & 1.00 (0.11) \\
 &  & U-FMR & 0.006 (0.002) & 0.005 (0.004) & 16.2 (17.4) & 0.56 (0.10) & 0.011 (0.020) & 0.001 (0.001) & 74.0 (26.2) & 0.68 (0.06) \\
 & \multirow{2}{*}{C} & MV-FMR & 0.002 (0.002) & 0.000 (0.001) & 86.1 (18.1) & 0.89 (0.07) & 0.005 (0.003) & 0.003 (0.003) & 50.8 (11.1) & 0.83 (0.04) \\
 &  & U-FMR & 0.013 (0.022) & 0.002 (0.001) & 69.9 (28.2) & 0.63 (0.08) & 0.006 (0.002) & 0.005 (0.004) & 15.4 (17.5) & 0.55 (0.09) \\
 & \multirow{2}{*}{D} & MV-FMR & 0.001 (0.000) & 0.000 (0.001) & 72.0 (13.3) & 0.99 (0.09) & 0.001 (0.001) & 0.001 (0.001) & 67.4 (16.0) & 0.92 (0.07) \\
 &  & U-FMR & 0.001 (0.001) & 0.001 (0.002) & 37.2 (21.0) & 0.61 (0.09) & 0.001 (0.000) & 0.001 (0.002) & 36.7 (19.4) & 0.62 (0.12) \\
\midrule
\multirow{8}{*}{3} & \multirow{2}{*}{A} & MV-FMR & 0.007 (0.005) & 0.003 (0.003) & 57.2 (12.6) & 0.76 (0.05) & 0.005 (0.003) & 0.002 (0.002) & 63.8 (11.8) & 0.93 (0.06) \\
 &  & U-FMR & 0.044 (0.042) & 0.028 (0.016) & 30.9 (25.3) & 0.58 (0.06) & 0.020 (0.050) & 0.007 (0.005) & 44.0 (27.5) & 0.51 (0.10) \\
 & \multirow{2}{*}{B} & MV-FMR & 0.005 (0.003) & 0.003 (0.003) & 48.8 (11.0) & 0.77 (0.07) & 0.001 (0.001) & 0.000 (0.001) & 90.1 (12.7) & 1.12 (0.09) \\
 &  & U-FMR & 0.008 (0.002) & 0.007 (0.006) & 11.4 (14.1) & 0.54 (0.10) & 0.017 (0.064) & 0.002 (0.002) & 60.8 (26.8) & 0.46 (0.11) \\
 & \multirow{2}{*}{C} & MV-FMR & 0.003 (0.002) & 0.001 (0.001) & 82.5 (19.2) & 0.82 (0.07) & 0.004 (0.002) & 0.003 (0.002) & 56.3 (11.3) & 0.90 (0.06) \\
 &  & U-FMR & 0.033 (0.039) & 0.017 (0.013) & 41.8 (25.4) & 0.54 (0.05) & 0.005 (0.001) & 0.004 (0.004) & 17.2 (18.1) & 0.61 (0.11) \\
 & \multirow{2}{*}{D} & MV-FMR & 0.001 (0.001) & 0.000 (0.001) & 63.5 (16.8) & 0.73 (0.05) & 0.001 (0.000) & 0.000 (0.001) & 68.3 (14.5) & 0.99 (0.08) \\
 &  & U-FMR & 0.002 (0.001) & 0.002 (0.002) & 17.8 (20.0) & 0.60 (0.09) & 0.001 (0.000) & 0.001 (0.001) & 39.0 (23.0) & 0.58 (0.10) \\
\bottomrule
\end{tabular}%
}
\vspace{0.5em}
\caption{MV-FMR and U-FMR performance under alternative functional forms of the exposure-outcome relationship, Cases A--D, for Scenarios 1 and 3. Continuous outcome. ISE: Integrated Squared Error (mean (SD) across time points); ISB: Integrated Squared Bias (mean (SD) across time points); CO: pointwise 95\% bootstrap CI coverage rate, \%; SEr: mean SE calibration ratio (mean (SD) across time points).}
\label{supp:tab:shapes_sensitivity_comprehensive}
\end{table}

\renewcommand{\thetable}{S4}
\setcounter{table}{0}
\section*{Supplementary Table 4: Sensitivity analysis of MV-FMR performance (coverage) under alternative overlapping instruments, mediation strength, and longitudinal sparsity for the continuous outcome}
\begin{table}[h]
\centering
\small
\setlength{\tabcolsep}{7pt}
\renewcommand{\arraystretch}{1.18}
\begin{tabular}{@{}lcccc@{}}
\toprule

\multicolumn{5}{@{}l}{A. Overlapping instruments, $p_{12}$, Scenario 1} \\
\addlinespace[0.15em]
$p_{12}$ 
& \textbf{0.15} & 0.30 & 0.50 & 0.80 \\
\cmidrule(lr){1-5}
$\beta_1(t)$ 
& 85.4 (16.4) & 84.8 (19.0) & 84.6 (20.0) & 81.5 (22.1) \\
$\beta_2(t)$ 
& 86.5 (15.2) & 85.9 (14.9) & 85.7 (17.8) & 83.2 (19.8) \\

\addlinespace[0.25em]
\midrule
\addlinespace[0.15em]

\multicolumn{5}{@{}l}{B. Mediation strength, $\gamma$, Scenario 3} \\
\addlinespace[0.15em]
$\gamma$ 
& 0.15 & \textbf{0.30} & 0.50 & 0.80 \\
\cmidrule(lr){1-5}
$\beta_1(t)$ 
& 60.8 (25.7) & 57.9 (25.9) & 57.7 (27.0) & 53.7 (28.5) \\
$\beta_2(t)$ 
& 62.3 (24.5) & 62.6 (25.2) & 66.5 (24.0) & 70.5 (24.1) \\

\addlinespace[0.25em]
\midrule
\addlinespace[0.15em]

\multicolumn{5}{@{}l}{C. Longitudinal sparsity, $n_{\text{sparse}}$, Scenario 1} \\
\addlinespace[0.15em]
$n_{\text{sparse}}$ 
& 5 & \textbf{10} & 25 & 50 \\
\cmidrule(lr){1-5}
$\beta_1(t)$ 
& 88.0 (17.6) & 85.4 (16.4) & 70.8 (13.2) & 58.8 (12.5) \\
$\beta_2(t)$ 
& 88.2 (17.5) & 86.5 (15.2) & 73.5 (12.6) & 57.7 (11.6) \\
\bottomrule
\end{tabular}
\vspace{0.5em}
\caption{Coverage results for MV-FMR under different degrees of instrument overlap, mediation strength, and longitudinal data sparsity for the continuous outcome. Values are pointwise 95\% bootstrap CI coverage rates, reported as mean (SD), for $\beta_1(t)$ and $\beta_2(t)$. Bold values indicate the reference setting used in the main simulation scenarios.}
\label{supp:tab:sensitivity_coverage_continuous}
\end{table}

\renewcommand{\thetable}{S5}
\setcounter{table}{0}
\section*{Supplementary Table 5: Sensitivity analysis of MV-FMR performance under alternative overlapping instruments, mediation strength, and longitudinal sparsity for the binary outcome}
\begin{table}[H]
\centering
\small
\setlength{\tabcolsep}{7pt}
\renewcommand{\arraystretch}{1.18}
\begin{tabular}{@{}l l cccc@{}}
\toprule

\multicolumn{6}{@{}l}{A. Overlapping instruments, $p_{12}$, Scenario 1} \\
\addlinespace[0.15em]
 & & \textbf{0.15} & 0.30 & 0.50 & 0.80 \\
\cmidrule(lr){3-6}
\multirow{2}{*}{ISE} 
 & $\beta_1(t)$ & 0.029 (0.009) & 0.031 (0.007) & 0.028 (0.008) & 0.033 (0.038) \\
 & $\beta_2(t)$ & 0.029 (0.009) & 0.031 (0.008) & 0.028 (0.007) & 0.029 (0.008) \\
\cmidrule(lr){1-6}
\multirow{2}{*}{CO (\%)} 
 & $\beta_1(t)$ & 32.6 (23.8) & 25.4 (18.7) & 32.6 (22.3) & 32.0 (23.5) \\
 & $\beta_2(t)$ & 32.2 (24.2) & 27.1 (18.4) & 30.8 (21.4) & 31.7 (21.9) \\

\addlinespace[0.25em]
\midrule
\addlinespace[0.15em]

\multicolumn{6}{@{}l}{B. Mediation strength, $\gamma$, Scenario 3} \\
\addlinespace[0.15em]
 & & 0.15 & \textbf{0.30} & 0.50 & 0.80 \\
\cmidrule(lr){3-6}
\multirow{2}{*}{ISE} 
 & $\beta_1(t)$ & 0.034 (0.010) & 0.042 (0.009) & 0.030 (0.010) & 0.042 (0.012) \\
 & $\beta_2(t)$ & 0.034 (0.010) & 0.041 (0.009) & 0.039 (0.010) & 0.042 (0.012) \\
\cmidrule(lr){1-6}
\multirow{2}{*}{CO (\%)} 
 & $\beta_1(t)$ & 31.0 (24.2) & 23.8 (20.0) & 18.9 (17.5) & 15.8 (16.5) \\
 & $\beta_2(t)$ & 31.3 (23.5) & 23.2 (19.4) & 18.5 (17.4) & 17.6 (14.6) \\

\addlinespace[0.25em]
\midrule
\addlinespace[0.15em]

\multicolumn{6}{@{}l}{C. Longitudinal sparsity, $n_{\text{sparse}}$, Scenario 1} \\
\addlinespace[0.15em]
 & & 5 & \textbf{10} & 25 & 50 \\
\cmidrule(lr){3-6}
\multirow{2}{*}{ISE} 
 & $\beta_1(t)$ & 0.100 (0.013) & 0.029 (0.009) & 0.004 (0.007) & 0.003 (0.003) \\
 & $\beta_2(t)$ & 0.099 (0.012) & 0.029 (0.009) & 0.004 (0.011) & 0.003 (0.004) \\
\cmidrule(lr){1-6}
\multirow{2}{*}{CO (\%)} 
 & $\beta_1(t)$ & 12.9 (14.3) & 32.6 (23.8) & 99.5 (2.6) & 99.9 (0.5) \\
 & $\beta_2(t)$ & 12.9 (16.4) & 32.2 (24.2) & 99.9 (0.4) & 99.9 (0.6) \\
\bottomrule
\end{tabular}
\vspace{0.5em}
\caption{Sensitivity of MV-FMR under different degrees of overlapping instruments, mediation strength, and longitudinal sparsity for the binary outcome. All metrics reported as mean (SD). ISE: Integrated Squared Error; CO (\%): pointwise 95\% bootstrap CI coverage rate. Results are reported separately for $\beta_1(t)$ and $\beta_2(t)$. Bold values indicate the reference setting used in the main simulation scenarios.}
\label{supp:tab:sensitivity_binary_ise_coverage}
\end{table}

{\color{black}
\renewcommand{\thetable}{S6}
\setcounter{table}{0}
\section*{Supplementary Table 6: Sensitivity analysis of MV-FMR performance under varying number of genetic instruments}
\begin{table}[H]
\centering
\resizebox{\textwidth}{!}{%
\begin{tabular}{cccccccccccc}
\toprule
 & & \multicolumn{3}{c}{$X_1$ (cFF)} & \multicolumn{3}{c}{$X_2$ (cFF)} & \multicolumn{2}{c}{$\beta_1(t)$} & \multicolumn{2}{c}{$\beta_2(t)$} \\
\cmidrule(lr){3-5}\cmidrule(lr){6-8}\cmidrule(lr){9-10}\cmidrule(lr){11-12}
$P$ & $h^2$ & PC1 & PC2 & PC3 & PC1 & PC2 & PC3 & ISE & CO (\%) & ISE & CO (\%) \\
\midrule
25 & $\approx 4\%$ & 8.48 (2.39) & 6.70 (1.84) & 1.92 (0.65) & 8.08 (2.28) & 7.08 (2.11) & 2.01 (0.60) & 0.011 (0.013) & 90.4 (16.2) & 0.016 (0.021) & 85.3 (20.5) \\
\textbf{50} & $\approx 7\%$ & 8.11 (1.48) & 6.99 (1.47) & 1.87 (0.37) & 8.06 (1.53) & 6.80 (1.43) & 1.87 (0.34) & 0.009 (0.011) & 85.4 (16.4) & 0.008 (0.009) & 86.5 (15.2) \\
75 & $\approx 10\%$ & 8.07 (1.28) & 6.63 (1.13) & 1.80 (0.35) & 8.33 (1.31) & 6.92 (1.13) & 1.81 (0.32) & 0.009 (0.011) & 81.3 (17.3) & 0.008 (0.009) & 83.9 (17.9) \\
100 & $\approx 13\%$ & 8.46 (1.14) & 6.49 (0.85) & 1.74 (0.25) & 8.35 (1.05) & 6.49 (0.88) & 1.74 (0.23) & 0.010 (0.010) & 78.8 (16.8) & 0.008 (0.007) & 82.6 (13.1) \\
\bottomrule
\end{tabular}%
}
\vspace{0.5em}
\caption{Sensitivity of MV-FMR performance to the number of genetic instruments $P$ (Scenario~1, pleiotropy, linear effects, continuous outcome). $h^2$: time-averaged analytical heritability (Supplementary Method~4). cFF per PC: conditional F-statistic for each retained principal component, reported as mean (SD) across replications. All metrics reported as mean (SD). ISE: Integrated Squared Error; ISB: Integrated Squared Bias; CO (\%): pointwise 95\% bootstrap CI coverage rate. Bold values indicate the reference setting used in the main simulation scenarios.}
\label{supp:tab:sensitivity_J}
\end{table}
}

{\color{black}
\newpage
\renewcommand{\thetable}{S7}
\setcounter{table}{0}
\section*{Supplementary Table 7: Sensitivity analysis of MV-FMR performance under varying number of exposures}
\begin{table}[ht]
\centering
\label{supp:tab:supp_m_exp}
\begin{tabular}{clcccc}
\toprule
$m$ & Exposure & ISE & ISB & CO (\%) & SEr \\
\midrule
  \multirow{2}{*}{$\textbf{2}$} & $\beta_{1}$ & 0.009 (0.011) & 0.002 (0.001) & 85.4 (16.4) & 0.96 (0.08) \\
   & $\beta_{2}$ & 0.008 (0.009) & 0.002 (0.001) & 86.5 (15.2) & 1.07 (0.07) \\
  \midrule
  \multirow{3}{*}{$3$} & $\beta_{1}$ & 0.011 (0.013) & 0.002 (0.001) & 85.8 (14.1) & 1.01 (0.07) \\
   & $\beta_{2}$ & 0.010 (0.018) & 0.001 (0.001) & 86.8 (15.7) & 0.96 (0.05) \\
   & $\beta_{3}$ & 0.009 (0.009) & 0.001 (0.001) & 86.9 (13.9) & 1.04 (0.07) \\
  \midrule
  \multirow{4}{*}{$4$} & $\beta_{1}$ & 0.012 (0.016) & 0.002 (0.001) & 90.1 (12.1) & 1.11 (0.07) \\
   & $\beta_{2}$ & 0.014 (0.024) & 0.001 (0.001) & 89.0 (16.7) & 0.97 (0.07) \\
   & $\beta_{3}$ & 0.011 (0.012) & 0.001 (0.001) & 90.7 (10.9) & 1.08 (0.06) \\
   & $\beta_{4}$ & 0.011 (0.012) & 0.001 (0.001) & 92.0 (9.3) & 1.13 (0.06) \\
  \midrule
  \multirow{5}{*}{$5$} & $\beta_{1}$ & 0.016 (0.016) & 0.002 (0.001) & 93.3 (6.1) & 1.14 (0.08) \\
   & $\beta_{2}$ & 0.018 (0.039) & 0.001 (0.001) & 92.6 (12.7) & 1.01 (0.07) \\
   & $\beta_{3}$ & 0.015 (0.017) & 0.001 (0.001) & 93.1 (12.8) & 1.10 (0.06) \\
   & $\beta_{4}$ & 0.013 (0.014) & 0.001 (0.001) & 93.6 (8.6) & 1.19 (0.07) \\
   & $\beta_{5}$ & 0.019 (0.025) & 0.001 (0.001) & 90.4 (15.3) & 1.03 (0.05) \\
\bottomrule
\end{tabular}
\caption{MV-FMR performance as the number of exposures $m$ increases (Scenario~1, pleiotropy, linear effect for all the exposures, $N=5000$, $P=50$, $n_{\text{sparse}}=10$, $p_{12}=0.15$). All metrics reported as mean (SD). ISE: Integrated Squared Error; ISB: Integrated Squared Bias across time points; CO (\%): pointwise 95\% bootstrap CI coverage rate; SEr: SE calibration ratio across time points. Bold values indicate the reference setting used in the main simulation scenarios.}
\end{table}
}

\newpage
\section*{Supplementary Figure 1: Mean time-varying causal effect estimates across different exposure-outcome model combinations with a binary outcome}
\begin{figure}[H]
    \centering
\includegraphics[width=0.9\textwidth,height=0.72\textheight,keepaspectratio]{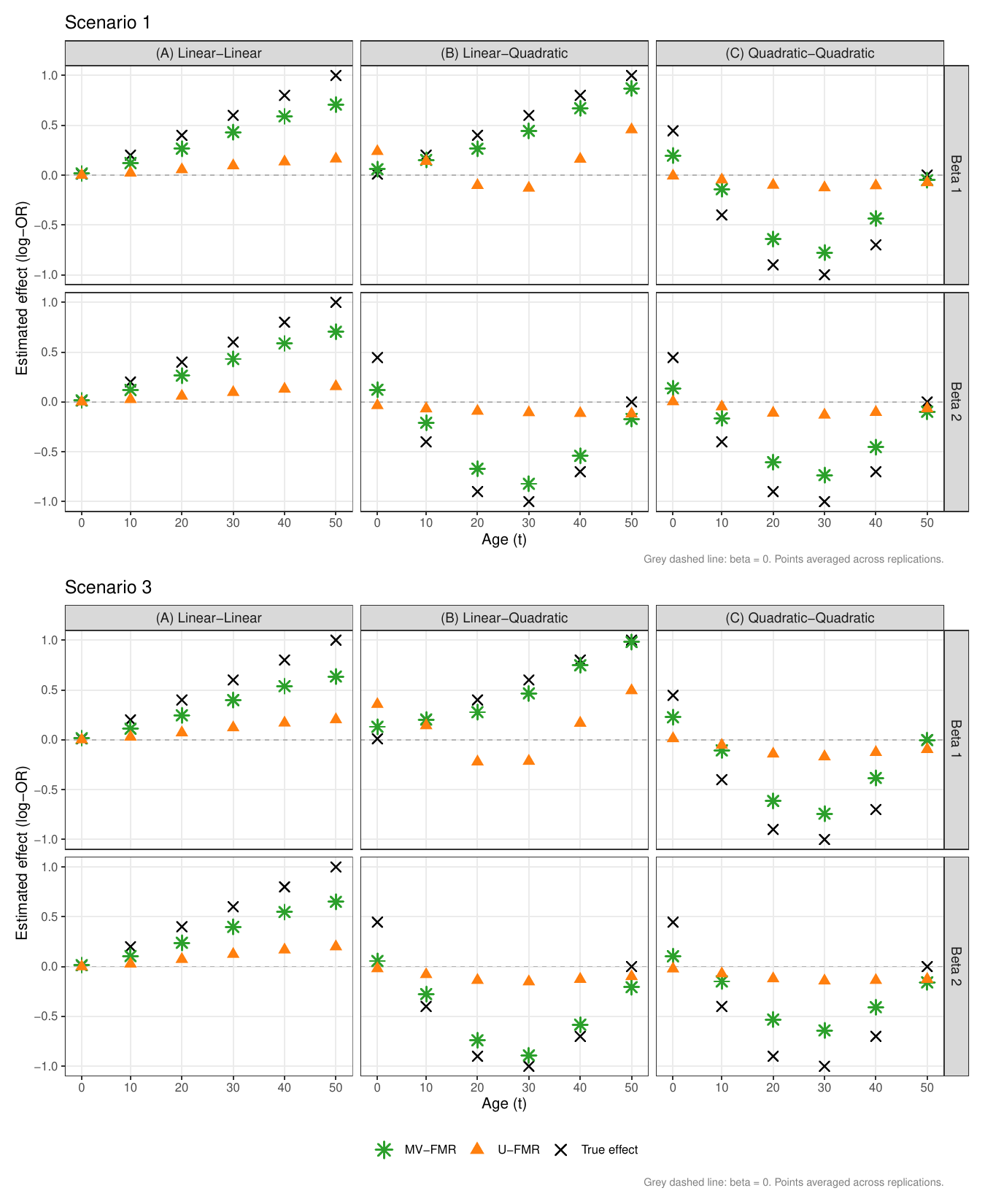}
    \caption{Mean time-varying causal effect estimates at six time points for comparing MV-FMR and U-FMR with binary outcome. Columns represent different exposure-outcome model combinations; rows show $\beta_1(t)$ and $\beta_2(t)$. Black crosses indicate the true causal effects; green stars show MV-FMR estimates; orange triangles show U-FMR estimates (included as a benchmark). Top panel: Scenario~1. Bottom panel: Scenario~3.}
    \label{supp:fig:sim_combined_beta_binary}
\end{figure}

{\color{black}
\newpage
\section*{Supplementary Figure 2: Mean pointwise integrated squared error at six representative time points (continuous outcome)}
\begin{figure}[H]
    \centering
    \includegraphics[width=\textwidth]{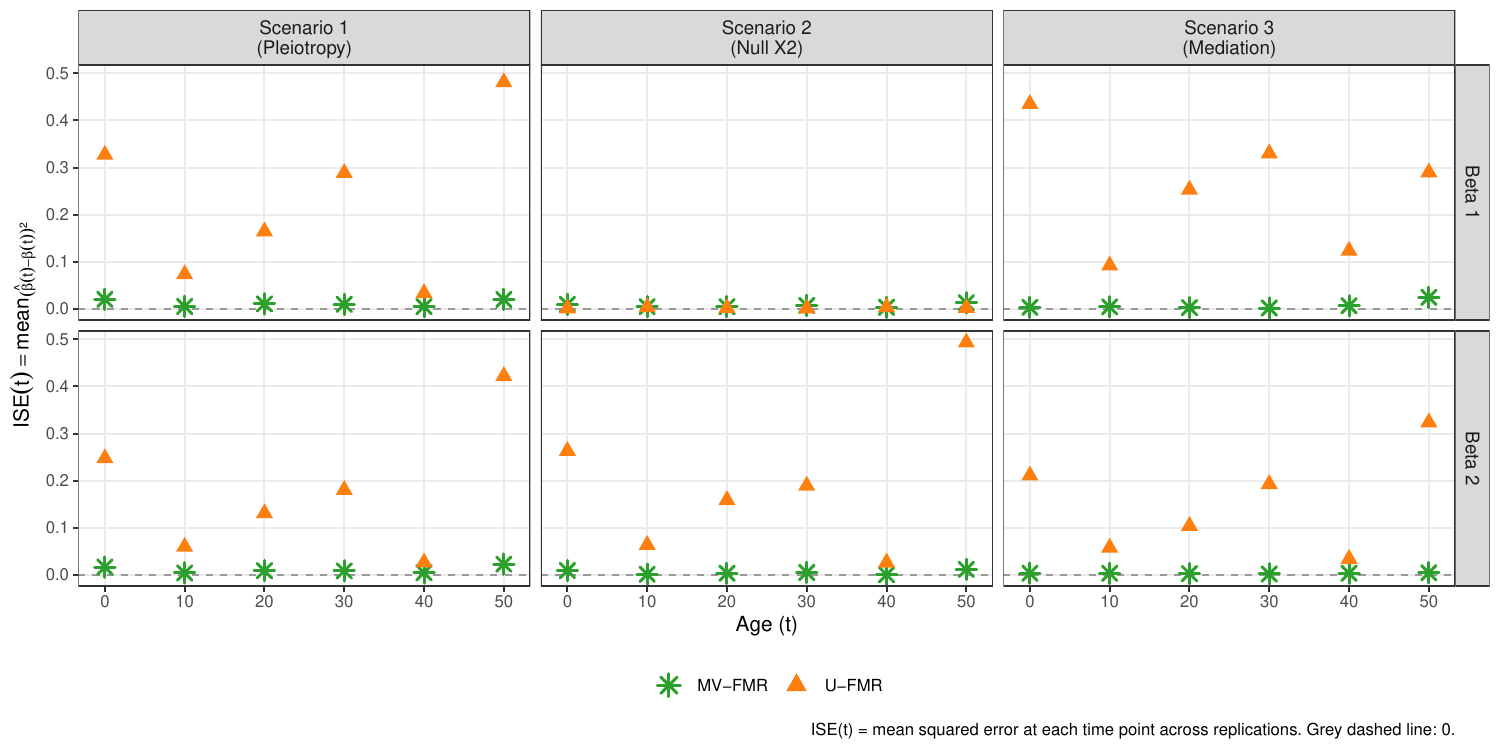}
    \caption{Mean pointwise Integrated Squared Error (ISE) at six representative time points (0, 10, 20, 30, 40, 50), averaged across 500 simulation replications, for a continuous outcome. Rows show $\beta_1(t)$ and $\beta_2(t)$; columns show Scenario~1 (pleiotropy), Scenario~2 (null $X_2$), and Scenario~3 (mediation). Green stars indicate MV-FMR; orange triangles indicate U-FMR (benchmark). The integrated value of these curves corresponds to the ISE reported in Table~1.}
    \label{supp:fig:ise_profile}
\end{figure}
}

{\color{black}
\section*{Supplementary Figure 3: Pointwise integrated squared bias at six representative time points (continuous outcome)}
\begin{figure}[H]
    \centering
    \includegraphics[width=\textwidth]{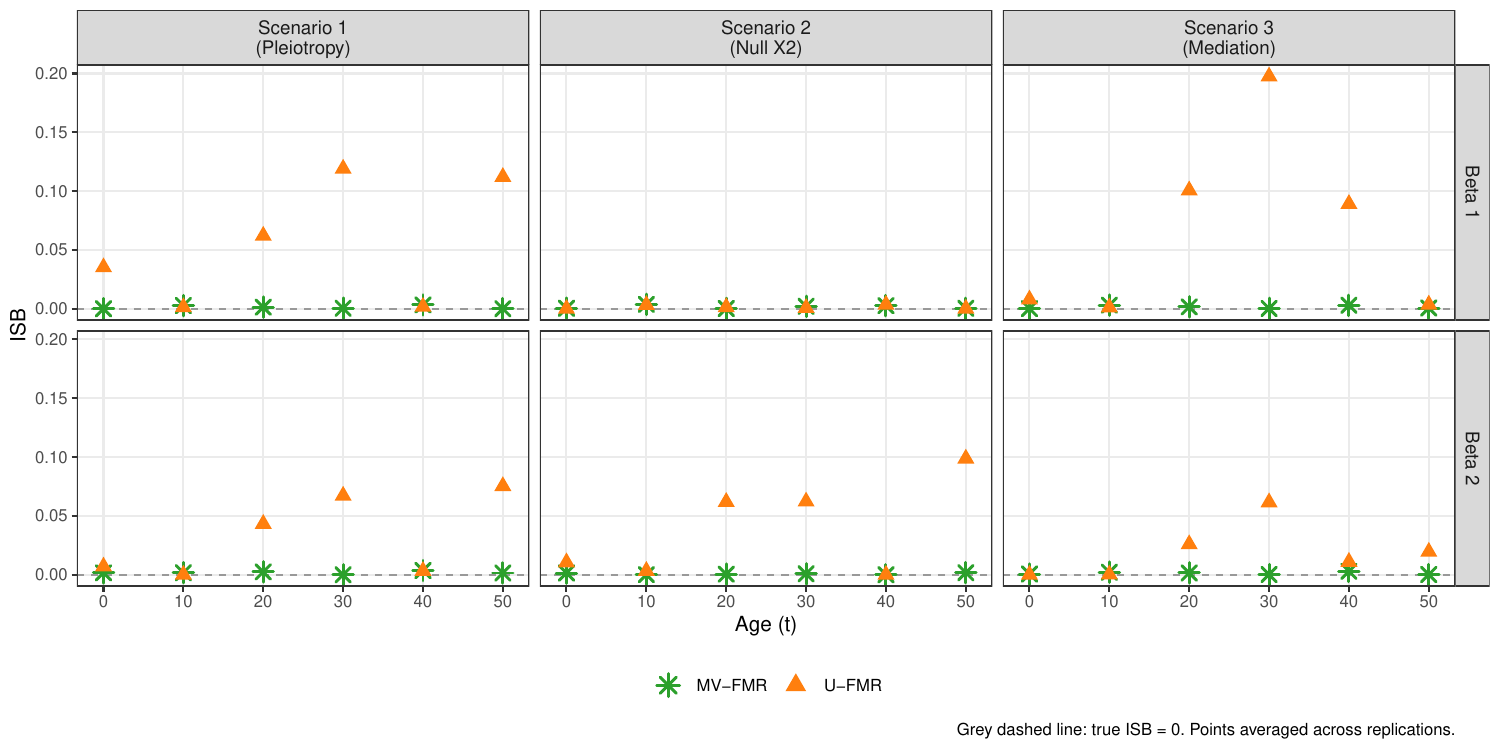}
    \caption{Pointwise Integrated Squared Bias (ISB) at six representative time points (0, 10, 20, 30, 40, 50), for a continuous outcome. Rows show $\beta_1(t)$ and $\beta_2(t)$; columns show Scenario~1 (pleiotropy), Scenario~2 (null $X_2$), and Scenario~3 (mediation). Green stars indicate MV-FMR; orange triangles indicate U-FMR (benchmark). The grey dashed line at zero is the reference value for unbiased estimators. The mean value of these estimates corresponds to the ISB reported in Table~1.}
    \label{supp:fig:isb_profile}
\end{figure}
}

{\color{black}
\newpage
\section*{Supplementary Figure 4: Distribution of pooled SBP and BMI measurement ages among UK Biobank participants (ages 40--80)}
\begin{figure}[H]
    \centering
    \includegraphics[width=0.95\textwidth]{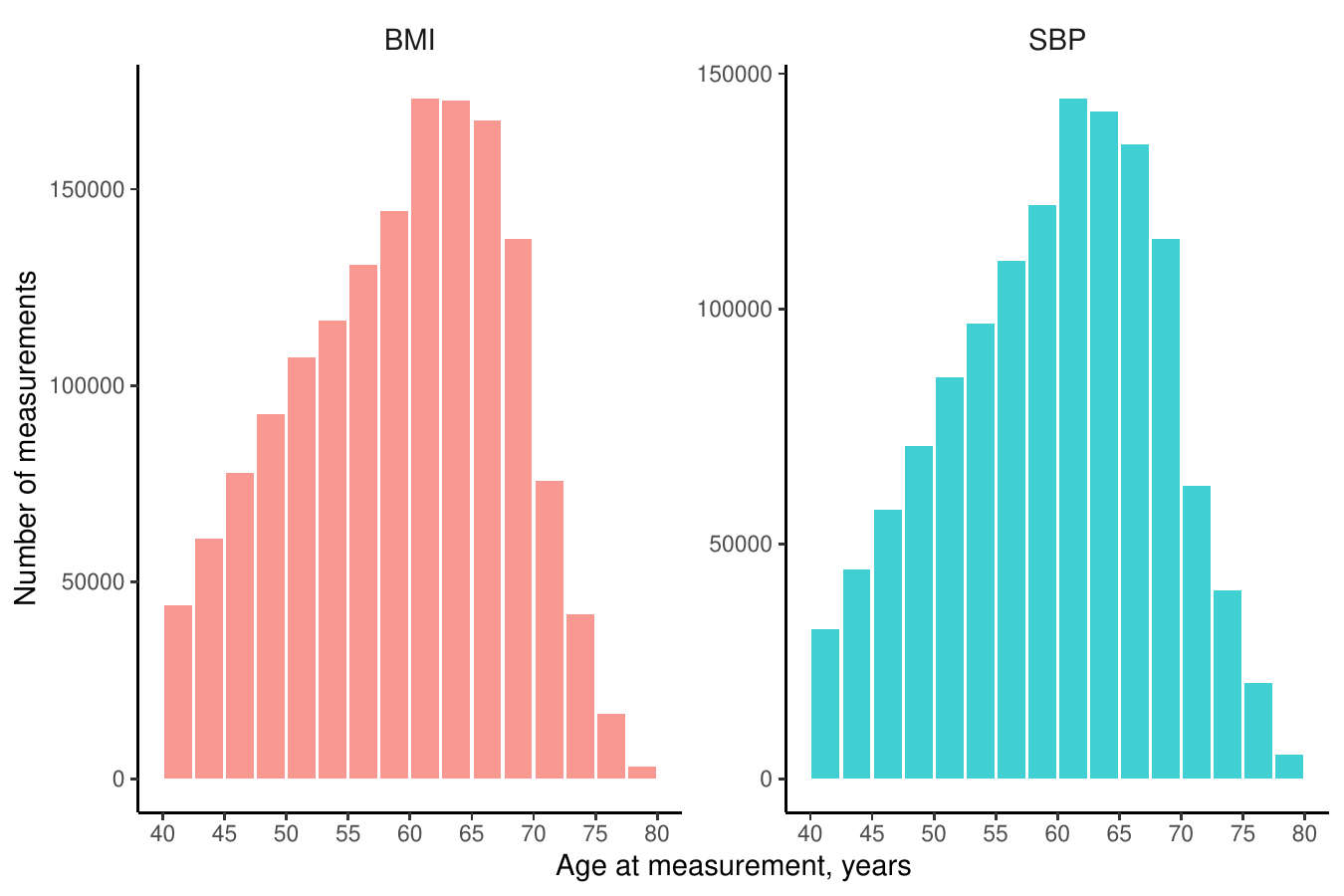}
    \caption{Distribution of pooled SBP and BMI measurement ages among UK Biobank European-ancestry participants with genetic data across the full 40--80 year age range. Each bar represents the number of measurements pooled across individuals within each 2.5-year age bin. Measurements originate from primary care records and UK Biobank assessment visits.}
    \label{supp:fig:measurement_dist_40_80}
\end{figure}

\newpage
\section*{Supplementary Figure 5: Distribution of pooled SBP and BMI measurement ages within the analysis cohort (ages 50--70)}
\begin{figure}[H]
    \centering
    \includegraphics[width=0.95\textwidth]{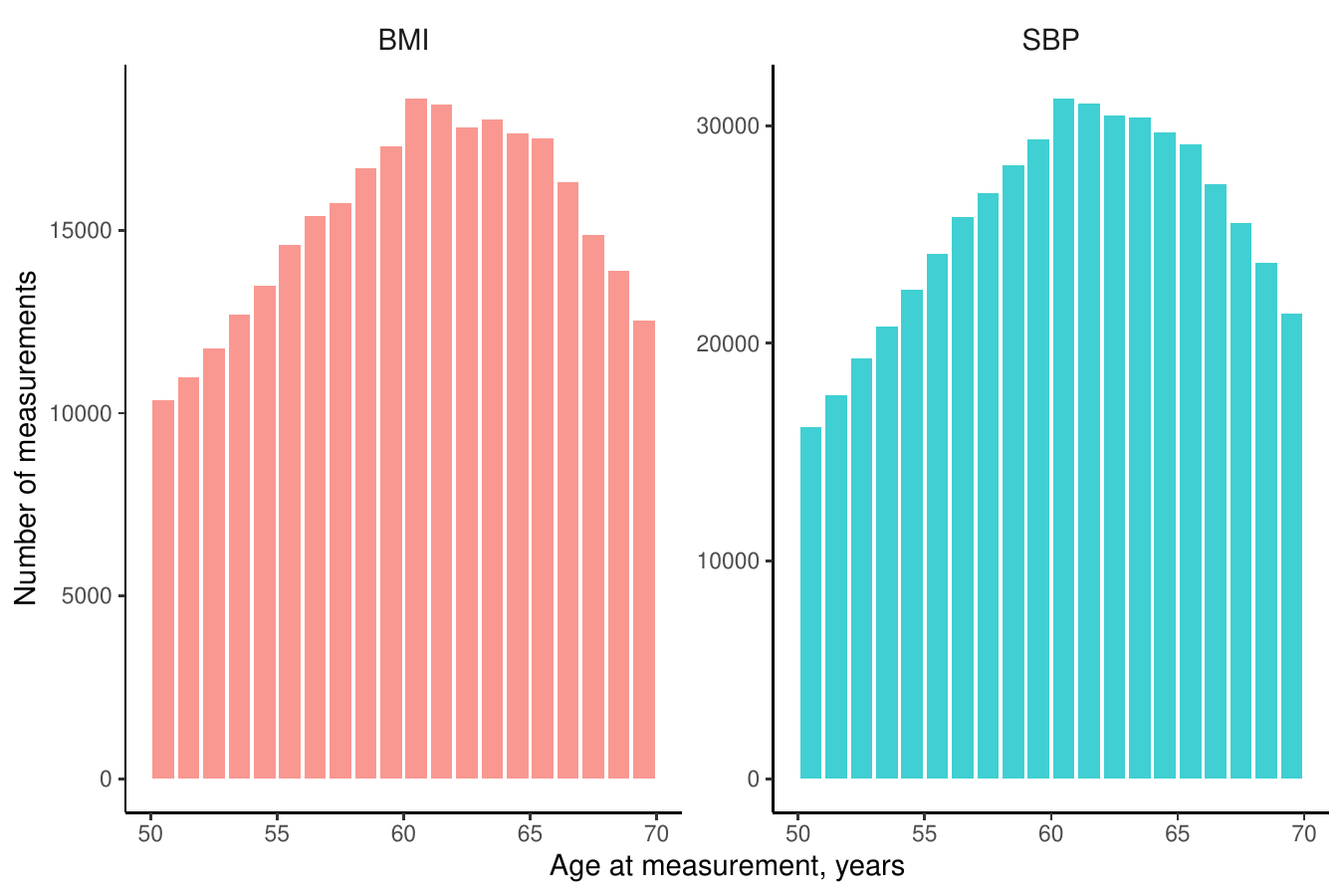}
    \caption{Distribution of pooled SBP and BMI measurement ages within the selected cohort of 59,907 UK Biobank participants, restricted to the 50--70 year window. Each bar represents the number of measurements pooled across individuals within each one-year age bin. Although individual-level trajectories were sparse, pooling across individuals provides dense and approximately uniform temporal coverage of the analysis domain.}
    \label{supp:fig:measurement_dist_50_70}
\end{figure}
}

\end{document}